\documentstyle[epsfig]{lmamult}
\begin{document}

\title{Slow Dynamics and Aging in Spin Glasses}

\author{Eric Vincent, Jacques Hammann, Miguel Ocio,
Jean-Philippe Bouchaud and Leticia F. Cugliandolo}
\authorrunning{E.Vincent, J.Hammann, M.Ocio, J.P.Bouchaud, L.Cugliandolo}

\institute{Service de Physique de l'Etat Condens\'e, CEA Saclay, 91191
 Gif-sur-Yvette Cedex, France }

\maketitle

\section{Introduction}

A crucial feature of the behavior of {\em real} spin glasses is the existence
of
{\it extremely  slow relaxational processes}.
Any field change causes a very long-lasting relaxation of the
magnetization and, the response to an ac excitation
is noticeably delayed.  In addition, the characteristics of this slow
dynamics evolve during the time spent in the spin-glass phase: the systems {\em
age}.

Aging effects in real spin glasses have been layed down by
experiments
\cite{lundgren1983,aging1,aging2}
at a time where there was already an intense theoretical activity on the
equilibrium properties of mean-field spin-glass models \cite{beyond}.
Experimentalists started comprehensive studies of the non-equilibrium dynamics,
which
happened to bring very instructive surprises, while in the meantime
theoreticians developed extremely sophisticated
methods for progressing towards solutions of the equilibrium mean-field
problem, thence inventing incentive tools for
the statistical mechanics of disordered systems.

This early epoch was not the time for the most productive dialogue between both
parts.
The situation is very different now; experiment and theory have had, during
these last years, a fruitful interplay.
On the one hand, the problem of the non-equilibrium dynamics has now been
theoretically
addressed from very different points of view; scaling theories of domain growth
\cite{Brmo,FH,KH,clusterocio}, a phase-space approach motivated by the Parisi
solution to mean-field models \cite{viktor},
a percolation like picture in phase space \cite{sibani1},
random walks in phase space \cite{sibani2,jpbtrap,jpbdd},
mean-field treatments of some
simplified situations
\cite{Cuku,Cuku3},
are now providing us with various (and sometimes contradictory) lightings of
the experimental results,
together with
impulsing a thrilling debate on the sound nature of the spin-glass phase. On
the other hand, more
and more complex
experimental procedures \cite{jphys+,uppsaladT,ucladT} have been conceived
with the
aim of evidencing the
materialization of some abstract theoretical notions, like {\it e.g.}  the
ultrametric organization of states
or  the chaotic
dependence of the spin-spin correlation function on temperature.

In this paper, we recall some important experimental features of the spin glass
dynamics. Since we intend to picture some aspects of the present state of the
dialogue
between experimentalists and theoreticians,
we give a detailed description of several ways of scaling the data and of the
connection between these scalings and
the theoretical predictions. We obviously
give up any pretention of giving an exhaustive comparison of theory and
experiment; we  mainly focus here on a perspective
of spin glasses which proceeds from mean-field results \cite{Cuku}
(abundant discussions of the scaling theories can be found in
the literature of  the past few years).
In several occasions, we use as a guideline for the description of the
experimental results a probabilistic model that
views aging as a thermally activated random walk in a set of traps with a wide
distribution of trapping times \cite{jpbtrap}.

The slow dynamics of spin glasses - and, as well, of
 structural glasses \cite{angell} and other disordered systems - has been often
interpreted in terms of thermal activation over
barriers. One likes to think of a complex free-energy landscape (due to
frustration)  with peaks and valleys of all sizes. This picture has been
extensively used in the litterature; Refs. \cite{viktor,sibani2,jpbtrap,jpbdd,
ucladT,ray}
are examples of different ways of drawing conclusions from it.
In fact, the experiments never directly probe  free-energy valleys or
mountains,
but rather give access to relaxation rates
at various time scales, which may then be interpreted in terms of thermal
activation over free-energy barriers \cite{ucladT}.

However, when trying to describe the slow dynamics and aging of real
spin glasses with a
``phase-space" viewpoint, it is worth noticing that phase-space
is infinite dimensional irrespectively of the finite or infinite
dimensionality of real space. The geometrical properties of the
infinite-dimensional phase space  may put at work a different
(non-Arrhenius) mechanism for slow dynamics  that leads to
slowing-down and aging even in the absence of
metastable states \cite{Kula}.
The particle point in phase space, representing the system configuration,
slowly decays through almost flat regions.
This mechanism seems to be the one acting
in the dynamics of mean-field spin-glass models  (with a single aging
correlation scale)  \cite{Cuku} as well as in domain growth. In these models
one does not see neither a  severe change of behavior when approaching the
zero-temperature limit nor
rapid barrier-crossings from trap to trap
in numerical simulations \cite{Cuku3,Kula}.

We describe below some mean-field predictions which compare
rather well with the experiments at constant temperature. Whether one can
describe more subtle experimental results such as
temperature variation dependences, etc. with  mean-field models
and/or with the above non-Arrhenius phase
space geometrical description is still an open question.
The rather good agreement at constant temperature suggests that the
phase-space dynamical mechanism at the base of the dynamics of
spin glasses may be a combination of rapid activated processes and slow decay
through flat regions \cite{Cuku,Kula,jpbmm}.
We might thence be led to
revise our ``common sense understanding" of the slow dynamics in disordered
systems.

\vspace{2cm}

\section{Experimental Evidence for Non-Stationary Dynamics}

\subsection{Magnetization Relaxation in Response to a Field Change}

In a measurement of  the relaxation of the ``thermo-remanent magnetization"
(TRM),  the
system is cooled in a small field from above $T_g$ down to some $T_0<T_g$; it
then ``waits"
in the field at $T_0$ during a time $t_w$, after which the field is cut, and
the subsequent decrease
of the TRM from the field-cooled (FC) value is recorded as a function of $t$.
Following an ``immediate
fall-off" of the magnetization (depending on  the sample and on temperature, of
the order of 50 to 90 \%),
 a slow logarithmic-like relaxation takes place; it is believed to head towards
zero, although never reaching
an end at laboratory time scales.

These endless-like relaxation processes and, more crucially, 
 the existence of ``aging'' phenomena \cite{lundgren1983,aging1,aging2}
are  a salient feature of spin-glass dynamics: for different values of the
waiting time $t_w$, different TRM-decay curves are obtained, as is
evidenced in Fig. 1.a.


\begin{figure}
\centerline{\epsfxsize=10cm
\epsffile{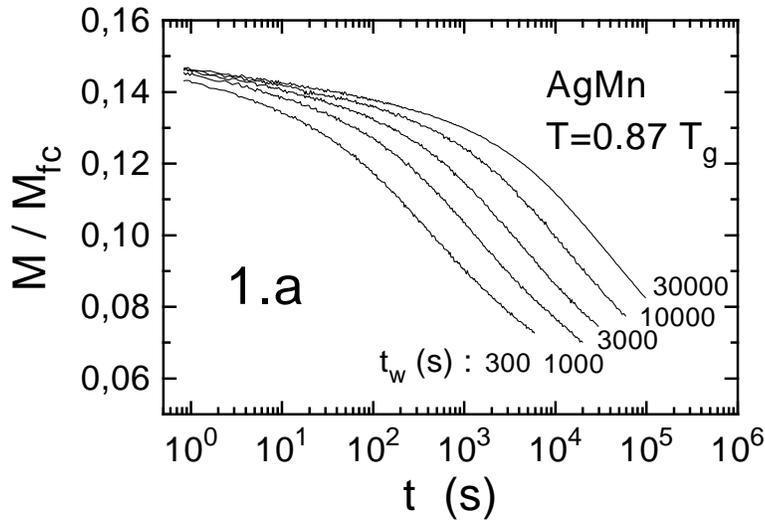}
}
\vspace{.2cm}
\caption{{\bf a.}
Thermo-remanent magnetization $M$, normalized by the field-cooled value
$M_{fc}$,
{\it vs.}~$t (s)$ ($\log_{10}$ scale) for the $Ag:Mn_{2.6\%}$ sample,
at $T=9 K=0.87 T_g$. The sample has been cooled in a $0.1\ Oe$ field from above
$T_g=10.4K$ to 9K; after waiting $t_w$, the field has been cut at $t=0$, and
the decaying magnetization recorded.
}
\end{figure}


The dynamics depends on two independent time-scales, $t$
(``observation time") and
$t_w$ (``waiting time"). This
dynamics is {\em  non-stationary}: the response at $t+t_w$ to an excitation at
$t_w$ depends on $t+t_w$
and $t_w$, and not only on
$t$ (breakdown of time-translational invariance). Qualitatively, one can see in
Fig. 1.a that the longer the
waiting time before cutting the field, the slower the overall response; the
initial fall-off is smaller, the relaxation
curve shows a slower decrease, the system has become ``stiffer''.
 Such aging
phenomena have been early
identified in the mechanical properties of glassy polymers
\footnote{A wide class of materials like {\it e.g.} PVC, PS, Epoxy, or even
bitumen, Wood's metal, amorphous
sugar and cheese \cite{struik}.
}
 \cite{struik}; the slow strain following the
application of a stress has been
recognized to depend on the time spent in the glassy phase.

In spin glasses, the aging phenomena have initially been explored using the
mirror experimental procedure of
the TRM \cite{lundgren1983}, in which the sample is cooled in zero-field, and
after $t_w$ a small field is applied.
As far as the field remains low enough (usually, in the range 0.1-10 Oe), both
procedures  are equivalent; the relaxation of this ``zero-field cooled
magnetization" (ZFC) follows the same
$t\ and\ t_w$ dependence as the relaxation of the TRM \cite{lund=fc,ryan}. More
precisely, it has been shown in \cite{lund=fc} that, for all
$t,t_w$  values ({\it i.e.} all along the
measured relaxations for various $t_w$), the sum of the ZFC-magnetization plus
the TRM equals the field-cooled
value. This is simply {\em linearity} in the response, since this experimental
result shows that {\em the sum of the
responses to different excitations} is equal to {\em the response to the sum of
both excitations} (the response to a
constant field being the field-cooled magnetization). That linearity holds for
all $t,t_w$ tells us that the presence
of the (sufficiently small) field does not influence the aging process: waiting
$t_w$ in zero field and then applying
a field during $t$ (ZFC case) is equivalent, for the dynamics, to applying a
field during $t_w$ and then waiting $t$
in zero field (TRM case). The only role played by the field in this context is
to reveal the dynamic properties of the
system. A recent study of the effect on the dynamics of increasing field values
can be found in \cite{ray,deltaH,chu}.

In the semi-log plot of Fig. 1.a, each curve shows an inflection point, and one
first quantitative estimate of the
$t_w$-effect on the relaxation is that this inflection point is located around
$\log t\simeq \log t_w$. This fact
has been noticed and given a physical meaning by Lundgren {\it et al.}
\cite{lundgren1983}. The relaxations are
slower than exponential; they do not correspond to a single characteristic
response time $\tau$, but are likely
to be parametrized with the help of a wide distribution $g_{t_w}(\tau)$, which
is defined hereby:
\begin{equation}   \label{m=sum}
m_{t_w}(t) \equiv \frac{M(t+t_w,t_w)}{M_{fc}}  =\int_{\tau_0}^\infty
g_{t_w}(\tau)\exp(-{t\over\tau})d\tau
\end{equation}
where $\tau_0\simeq 10^{-12} sec$ is a microscopic attempt time.
$M(t+t_w,t_w)$ is the ZFC or TRM, depending on the experiment, and
$M_{fc}(t+t_w)$
is the
field-cooled value at time $t+t_w$. In the figures we abbreviate
$M(t+t_w,t_w)/M_{fc}(t+t_w)= M/M_{fc}$.
Lundgren {\it et al.} have pointed out that taking
the derivative of  (\ref{m=sum}) with respect to $\log t$ gives access to the
distribution $g_{t_w}(\tau)$, since
\begin{equation}  \label{dmdlogt}
{dm_{t_w}(t)\over d\log t}=-\int_{\tau_0}^\infty
g_{t_w}(\tau){t\over\tau}\exp(-{t\over\tau})d\tau
\approx g_{t_w}(\tau=t)\ ,
\end{equation}
a rough approximation reflecting the sharp character of
${t\over\tau}\exp(-{t\over\tau})$ around $t=\tau$
(to be considered on a logarithmic scale, which is actually the scale which is
suggested by the measurements).
The plot of the relaxation derivatives shows bell-like shapes, with a broad
maximum around $\log t =\log t_w$,
and pictures $g_{t_w}(\tau)$ for various $t_w$ \cite{lundgren1983}. Thus, in a
first approximation, the aging phenomenon can be described
 as a {\em logarithmic shift towards longer times} of a wide spectrum of
response times
\footnote{
Indeed, the spin-glass properties do not exactly depend on $t_w$, but rather on
the {\em total} elapsed time $t_w+t$ (Sect. 4.2); they are evolving during the
TRM measurement itself \cite{aging2,nousaging}. The physical interpretation
\cite{lundgren1983} of $g_{t_w}(\tau)$  therefore remains approximate.
}. This shift is of the order
of $\log t_w$, and therefore suggests that the dynamics be the same as a
function of $t/t_w$.

 Let us call ``full aging" the pure  $t/t_w$ scaling,
 that is not far from being the correct one,
as seen in Fig. 1.b where the data from
Fig. 1.a is presented versus $t/t_w$. Most of the $t_w$-effect
has been accounted for, though some systematic departures remain, and are
worth being discussed (see Sect. 4.2).

\setcounter{figure}{0}
\begin{figure}
\centerline{\epsfxsize=10cm
\epsffile{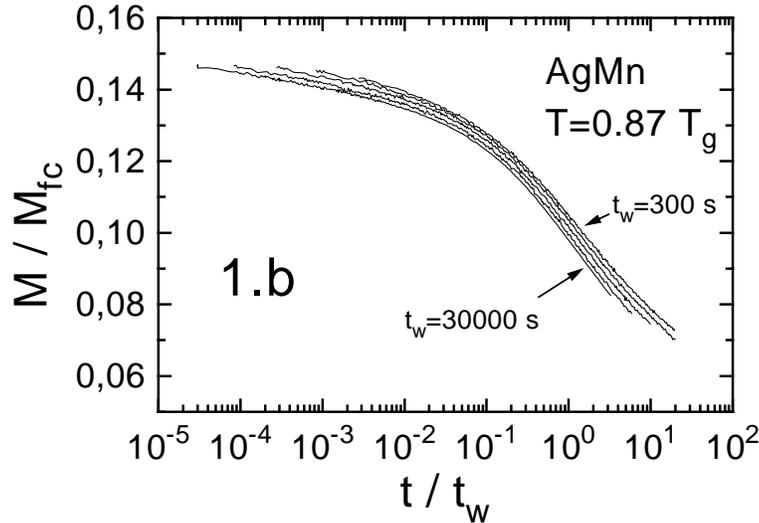}
}
\vspace{.2cm}
\caption{{\bf b.}
Same TRM data as in Fig. 1a,
 presented as a function of
$t/t_w$.
}
\end{figure}


\subsection{Ac susceptibility}

The approximate $t/t_w$ scaling of the TRM (or ZFC) curves is sufficient for a
description of  aging effects
 in ac experiments, where the in-phase and out-of-phase components of the
response to a small ac excitation
field at a frequency $\omega$ are measured. Aging is more visible (in relative
value) in the
out-of-phase component  $\chi''$ of the magnetic susceptibility, which
represents dissipation. The {\em observation
time}, corresponding to $t$ in TRM experiments, is here constant, equal to
$1/\omega$. When the
sample is cooled from above $T_g$ down to $T_0<T_g$, the susceptibility does
not immediately reach an
equilibrium value, but shows a slow relaxation as time goes on. We denote $t_a$
(``age") this time elapsed
from the quench into the spin-glass phase; in the TRM experiment, the
equivalent age is $t+t_w=t_a$.
The {\em non-stationary} character of the dynamics, which appears in the TRM
measurements as a dependence
on the two independent time scales $t$ and $t_w$, shows up in ac experiments as
a dependence of $\chi"$ on the
two variables  $\omega\ and\ t_a$.
This is clear in Fig. 2 where $\chi"$ at
various (low) frequencies $\omega$ is plotted as a function of $\omega t_a$;
applying a
vertical shift, the curves can all
be merged with respect to this reduced variable, which is equivalent to $t/t_w$
in TRM experiments.


\setcounter{figure}{1}
\begin{figure}
\centerline{\epsfxsize=9cm
\epsffile{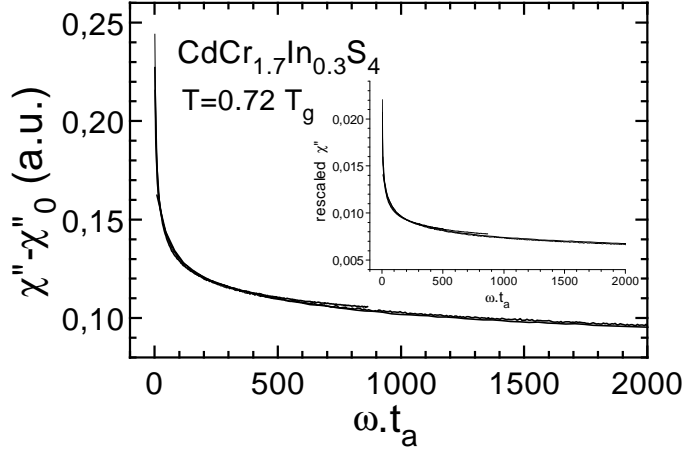}
}
\caption{
Out of phase susceptibility $\chi(\omega,t_a)$ {\it vs.}~$\omega \, t_a$ for
the insulating $CdCr_{1.7}In_{0.3}S_4$ sample. The four curves, corresponding
to
$\omega = 0.01, 0.03, 0.1, 1. \; Hz$, have been vertically shifted (see text).
$t_a$ is the total time elapsed from the quench (age). The inset shows a
scaling of the same data which follows from mean-field results (see Sect. 3.2).
}
\end{figure}

Thus, the approximate $t/t_w$ scaling obtained from TRM and ZFC experiments can
be fairly well
transposed to an $\omega.t_a$ scaling of the ac susceptibility
$\chi"(\omega.t_a)$. The vertical shift
corresponds to accounting for the various ``equilibrium values"
($\chi"_{eq}(\omega)=\lim
\chi"(\omega, t_a \to \infty$)) at different frequencies. In Fig. 2, for
technical reasons,
the zero of the scale has not been measured and the shift is arbitrary. As an
example of the relative orders of magnitude, let us
mention that, at $\omega=0.01Hz$, the amount of the relaxing part is roughly
equal to the equilibrium value; for $\chi'$ in the same conditions, it would be
of  the order of
10\% of the equilibrium value. The frequency-dependence of $\chi"_{eq}$ has
been determined in other
studies \cite{ryan,nousaging}; it can be represented by a power law with a very
small
exponent (or else a power law of a logarithm)
\begin{equation} \label{omalf}
\chi"_{eq}(\omega) \propto \omega^\alpha\ ,
\label{alpha}
\end{equation}
where $\alpha$ increases in the range $0.01-0.1$ when approaching $T_g$ from
below. This is
valid in the $10^{-2}-10^{5}Hz$ range which has been explored, and has been
measured rather in
insulating than in inter\-metallic spin glasses (due to eddy currents in
metals). Both classes of samples have been found to present the same general
spin-glass
behavior
\cite{aging2,nousaging}.
The $\omega.t_a$ scaling indicates that the smaller the frequency, the longer
the time $t_a$
 during which a significant relaxation, characteristic of aging effects, can be
found. Therefore,
 at higher frequencies ($\omega\ge 10 Hz$) aging disappears very rapidly,
yielding almost
instantaneously a stable value $\chi"_{eq}(\omega)$; conversely, at lower
frequencies, the
determination of $\chi"_{eq}(\omega)$ becomes problematic, implying
measurements over tens of hours or days.

\subsection{Time Regimes in ac and dc Experiments}

Let us summarize the conditions for observing either stationary or
non-stationary (aging) dynamics in ac($\chi"$) and dc (TRM or ZFC)
measurements. We can define two distinct time regimes, which apply to
both experiments.

\begin{itemize}
\item
For $\omega.t_a \to \infty$, ``equilibrium dynamics" is recovered in ac
experiments, in the
sense that only one time scale is needed: the dynamics is then {\em
stationary}. This time
regime corresponds, for TRM's, to $(t_w+t)/t=t_a/t \gg 1$ or equivalently $t\ll
t_w$, that
is the very beginning of the TRM-decay curves.
\vspace{.25cm}

\item
If $\omega t_a$ is comparable to $t_a$ ($\omega t_a = O(1)$)
in the $\chi''$ experiments
and, equivalently, $t$ is comparable to $t_w$ in the TRM experiments,
one observes non-stationary dynamics.
\newline
The $\chi''$ measurements are limited
to $\omega t_a > 1$ since $\omega.t_a < 1$ cannot
be experimentally realized (the harmonic response is not defined at
times shorter than one period). Hence, with $\chi''$ we can only explore the
beginning of the aging regime.
\newline
In contrast, aging is predominant in TRM-measurements
over the largest part
of the accessible time scale since, for TRM's,  $(\omega.t_a) \equiv
t_a/t=(t_w+t)/t$ becomes rapidly close
to 1 as the observation time $t$ elapses. Measuring the TRM decay we have
access to a larger time-window in the aging regime.
\newline
Therefore, aging in $\chi''$ can only be compared with the TRM decay at the
beginning of the aging regime --  that we call early epochs (also called
``quasi-stationary regime" in \cite{nousaging}).
This might explain why a full aging scaling seems to apply better to $\chi"$
results than to the TRM (see Fig. 1.b and 2).

\end{itemize}

\vspace{2cm}
\section{Aging Theories for Old Results}

\subsection{Scaling theories}

The different time scales in which {\em stationary \rm and \em non-stationary}
dynamics
are occurring are likely to be mapped onto length scales in the real space of
spins. When
approaching $T_g$ from above, the onset of a
critical regime has been characterized by the diverging behavior of
the characteristic time
in ac experiments and of the non-linear susceptibility in dc studies,
\cite{critical}. A thermodynamic phase transition would imply the divergence of
a characteristic correlation length $\xi$
 when $T \to T_g$. In this equilibrium picture
the spin-glass phase is believed to be an ensemble of
randomly
oriented spins, which are frozen due to infinite-range correlations
corresponding to a long-distance ``order".

However, aging shows
that indeed equilibrium has not been established when crossing $T_g$.
In several ``scaling
theories" \cite{Brmo,FH,KH,clusterocio} of non-equilibrium phenomena in spin
glasses, the spin
correlations are considered to be limited to some
{\em finite} range $\xi(t)$ (out-of-equilibrium situation); as time elapses,
spin rearrangements yield a slow
(due to frustration)
extension of equilibrium correlations,  towards the equilibrium situation
of infinite range
($t\to\infty,\xi(t)\to\infty$). An ac experiment at frequency $\omega$,
as well as a TRM or ZFC
relaxation at time $t$, can be viewed as probing the spin-glass excitations
at a given length
scale $L$ which should be an increasing function of the {\em characteristic
probe time}
$1/\omega$ for $\chi"$ and $t$ for TRM. In the ``droplet model" by Fisher and
Huse \cite{FH},
one has $L\propto \log^{1/\psi}t$ ($\psi\le d-1$), and in the ``domain model"
by Koper and
Hilhorst \cite{KH} $L\propto t^{p/d}$ ($p\sim 0.5$). These theories do not
aim at a
microscopic description at the scale of spins, but make this mapping of
{\em time} onto
{\em length} scales quantitative, in terms of scaling laws. At least
qualitatively,  these models provide us with a convenient picture of
aging phenomena, which is the following. For short probe times compared to the
age ($\omega.t_a\gg 1$, or $t/t_w\ll 1$), short-ranged excitations are
involved ($L\ll\xi(t)$), and the increase with time of $\xi(t)$ does not affect
 the dynamics, which is found to be stationary (no aging). Conversely, for
longer probe
times compared to the age ($\omega.t_a\sim 1$, or $t\ge t_w$), the
characteristic
length of the relevant excitations is of the same order of magnitude as
$\xi(t)$,
which is increasing due to aging, and the dynamic properties are strongly
affected
by aging (non-stationary dynamics).

 The accurate
quantitative
agreement with the data is still to be discussed (see {\it e.g.} Ref.
\cite{jpbdd}). Some
critical remarks to the simple scaling approaches, which rise
 up in view of other results (T-variation experiments), are discussed in Sect.
5.1.

\subsection{Non-Equilibrium Dynamics in Microscopic Theories}

\subsubsection{The Models.}

 The classical ``realistic'' microscopic model of spin glasses
is the 3-D Edwards-Anderson (3DEA) model \cite{Edan}
\begin{equation}  \label{H=}
H=-\sum_{\langle i ,j \rangle } J_{ij} S_i S_j
\; ,
\end{equation}
where $J_{ij}$ are Gaussian or bimodal random variables, $S_i$ are Ising spins
and  $\langle i,j \rangle$ represents a sum over first neighbours on
a cubic 3D lattice.  It is very difficult to obtain analytical
results for the statics or the  dynamics of 3DEA, in
consequence, even after more than  20 years  of research on the field
of spin glasses, very few results are available.
A lot of efforts have been devoted to the numerical study mainly of the
equilibrium properties of the 3DEA. Again, the situation is still pretty
unclear:
basic questions as to the existence of a thermodynamic phase transition
are still not answered \cite{3DEAsim}.

The standard mean-field extension of the 3DEA model is due to Sherrington and
Kirkpatrick (SK) \cite{Shki} and corresponds to the same interactions as in
(\ref{H=})
but with the sum extended to hold over all pairs of spins in the system.
The study of the SK model - and of some other related mean-field
 models - had for a long time been confined to the search for equilibrium
properties
\cite{beyond}.

\subsubsection{Numerical Simulations of Out-of-Equilibrum Phenomena.}

One may wonder whether aging, which did at first sight appear as some
imperfection
of the experiments, is really intrinsic to the Hamiltonian (\ref{H=}). We now
know that the answer is yes.
Only recently, attention has been paid to the
study of the out of equilibrium dynamics of microscopic spin-glass models.
Andersson {\it et al.} \cite{anders} and
Rieger \cite{rieger} reproduced in a numerical simulation the procedure of,
{\it e.g.},
the TRM experience using the 3DEA model. The results show that
it captures the main
features of real
spin glasses: both slow dynamics and aging effects.
Later, numerical simulations of the large $D$ ``hypercubic" spin-glass cell
in real space
showed that also
this model, that is expected to reproduce the SK model, when $D \to\infty$,
captures the main
characteristics of aging \cite{Cukuri}, thus confirming the previous analytical
results that we describe in the
following  paragraphs.

\subsubsection{Analytical Approach: Formulation and Definitions.}

Again, it only happened recently that
{\it analytical} developments evidenced aging effects in mean-field
spin-glass models \cite{Cuku}, showing that these simplified models can
describe,
at least qualitatively, the phenomenology of real spin glasses.
The idea in the case of mean-field spin-glass models is just to try to solve
the
{\it exact} dynamical equations derived for $N$, the number of dynamical
variables in the
model, tending to infinity. These equations are well-defined, have a unique
solution
 and, in the absence of a magnetic field, only involve the correlation and
the response functions (see Eqs.(\ref{autocorr}),(\ref{resp}) below for their
definitions).  The initial
 condition is chosen to
be random so as to mimic the initial configuration just after the quench
in the experimental situation. One then considers the large-time limits, but
only after having already
taken the thermodynamic limit $N\to\infty$ to obtain the asymptotic behavior
of the solution.

It is important to notice that in this approach it is not necessary to assume
{\it a priori} any
particular structure of phase space - typically to say that there are many
metastable
states due to frustration separated by high barriers - to obtain the dynamical
behavior of the problem. The solution can {\it a posteriori} be given a
geometrical interpretation
\cite{jpbtrap,jpbdd,Kula,jpbmm}.

The solution shows that the equations are self-consistently solved
in the large-time limit
by an {\it aging} solution. The reason why these equations can be solved is the
{\it weak long-term
memory} of the system \cite{Cuku}.
The results fall into the {\it weak-ergodicity breaking scenario} previously
proposed in \cite{jpbtrap,jpbdd} within the trap model (see Sect. 3.3
below). When looking at the
large-time dynamics of the system, the weak
long-term memory property allows us to neglect the contribution of any finite
time-interval after the
quenching time. The system forgets what happens in ``finite'' time intervals
with respect to the ``infinite''
observation time. It keeps, however, an averaged memory of its history.  The
weak-ergodicity breaking scenario
tells us that the evolution of the system
continues forever; the dynamics slows down as time elapses but the system is
never completely stopped
in its evolution. The waiting-time $t_w$ gives us an idea of the age of the
system.

In the following we shall be a bit more technical and describe the main
features of the formalism and the
solution.

The auto-correlation function is defined as
\begin{equation}
C(t+t_w,t_w) \equiv \frac{1}{N} \,
\sum_{i=1}^N \overline {\langle s_i(t+t_w) s_i(t_w) \rangle}
\; ,
\label{autocorr}
\end{equation}
with the overline representing a mean over different realizations of the
disorder and $\langle \, \rangle$
an average over different realizations of the thermal noise. We then define
\begin{equation}
 C_F(t) \equiv \lim_{t_w\rightarrow\infty} C(t+t_w,t_w)
 \;\;\;\;\;\;\;\;\;\;
 C_F(0) =1
\;\;\;\;\;\;\;\;\;\;
\lim_{t \rightarrow\infty} C_F(t) =  q_{EA} \ \ .
\label{limitc}
\end{equation}
This allows us to separate the auto-correlation into two {\it additive}
parts, a {\it stationary} term and an {\it aging} term $C_A$ \cite{Cuku,jpbdd}:
\begin{equation}
C(t+t_w,t_w) =
C_F(t) - q_{EA} + C_A(t+t_w,t_w) \ \ .
\end{equation}
In the absence of a magnetic field,
the weak ergodicity breaking scenario \cite{jpbtrap,jpbdd,Cuku} implies
\begin{equation} \label{Czero}
\lim_{t\rightarrow\infty} C(t+t_w,t_w) = 0 \;\;\;\;\; \forall \mbox{ fixed }
t_w
\; ,
\end{equation}
thus
\begin{eqnarray}
\begin{array}{rclrcl}
\lim_{t \to \infty} \lim_{tw \to \infty} C_A(t+t_w,t_w)
&=&
q_{EA}
\\
\lim_{t \to \infty} C_A(t+t_w,t_w) &=&
0 \; .
\end{array}
\end{eqnarray}
It will turn out that the two scales corresponding to these two limits are
well-separated for mean-field models,
in the sense that in the time-regime where $C_F$ varies then $C_A$ stays
constant,
and {\it viceversa}. In other words, one can think of $q_{EA}$ as a value of
the correlation separating
different ``correlation-scales'', $C>q_{EA}$ and $C<q_{EA}$: when $t \ll t_w$,
$C> q_{EA}$ and we have stationary
dynamics, while when $t \gg t_w$, $C< q_{EA}$ and we have non-stationary
dynamics and aging just as
 described in Sect. 2.3 for the general features of aging in spin glasses.
 \footnote{
In some numerical works, the form $C(t+t_w,t_w) = t^{-x(T)} \, \Phi(t/t_w)$ has
been often used
to scale the data for all times $t$
(\cite{rieger,Cukuri,Yo,Pariru}, see also \cite{nousaging} for a related
discussion of the experimental data). It should be remarked that, though at
first
glance
this scaling  seems to be similar to the one following from the WEB scenario,
it
implies quite a different conclusion for the global behavior of the system.
Note that if one takes
the limit $\lim_{t\to\infty}\lim_{t_w\to\infty}$, that corresponds to exploring
the {\it end of
the stationary dynamics}, Eqs.(\ref{limitc}) yield
$\lim_{t\to\infty}\lim_{t_w\to\infty}
C(t+t_w,t_w) = q_{EA}$, while $\lim_{t\to\infty}\lim_{t_w\to\infty} t^{-x(T)}
\,
\Phi(t/t_w) = \Phi(0) \, \lim_{t\to\infty} t^{-x(T)} = 0$.
The stationary dynamics {\it and} the fact that the correlation decays to
$q_{EA}$ at the end of this time-regime,
have a very clear geometrical interpretation \cite{jpbdd,Cuku,Kula}.
}

In the same way,
the response function can be equivalently separated into a {\it stationary} and
a {\it non-stationary} term
\begin{equation}
R(t+t_w,t_w)
\equiv
\frac{1}{N} \,
\sum_{i=1}^N
\left.
\frac{
\overline{
\partial \langle s_i(t+t_w) \rangle}
}{\delta h_i(t_w) }
\right|_{h=0}
=
R_F(t) + R_A(t+t_w,t_w)
\label{resp}
\; .
\end{equation}
with
\begin{eqnarray} \label{RFA}
R_F(t) &\equiv&  \lim_{tw\to\infty \; C(t+t_w,t_w)  > q_{EA}} R(t+t_w,t_w)
\quad and
\\
R_A(t+t_w,t_w) &\equiv&
 \lim_{tw\to\infty \; C(t+t_w,t_w)  < q_{EA}} R(t+t_w,t_w) \quad .
\end{eqnarray}

 $R_F(t)$ satisfies the fluctuation-dissipation theorem (FDT) and
$R_A(t+t_w,t_w)$ satisfies a
generalized FDT \cite{Cuku}
\begin{equation}
R_F(t) = - \frac{1}{T} \frac{d C_F(t)}{d t}
\;\;\;\;\;\;
R_A(t+t_w,t_w) =
\frac{X[C_A(t+t_w,t_w)]}{T} \;
\frac{ \partial C_A(t+t_w,t') }{\partial t'} |_{t'=tw}
\; ,
\label{X}
\end{equation}
with $0 \leq X[C_A] \leq 1$ a
monotonically increasing function of $0 \leq C_A \leq q_{EA}$.
($X=1$ corresponds to the usual FDT.)

With these definitions one can solve the asymptotic (large $t_w$)
dynamics of several mean-field disordered models \cite{Cuku,Cuku3,Frme,Cukule,
Cudd}.

\subsubsection{Analytical Approach: Stationary Regime.}

For all these models,
when $t$ is large,
the {\it stationary} part of the correlation function $C_F(t)$
decays with a power law
\begin{equation}
C_F(t)
\sim
q_{EA} + c_\alpha \left( \frac{\tau_0}{t} \right)^\alpha
\; ,
\label{FDTdecay}
\end{equation}
where $\tau_0$ is a microscopic time-scale and
$\alpha$ has precisely the same meaning as the exponent in (\ref{alpha}),
describing the frequency
dependence of the equilibrium out of phase susceptibility.

The temperature dependence of $\alpha$ depends on the model.
For models
\footnote{
Though the aim of Refs. \cite{Crhoso,Kiho} was to study the
equilibrium dynamics \`a la Sompolinsky \cite{So} - a different situation from
the out of
equilibrium occuring in experiments - the calculation of the
exponent $\alpha$ obtained in these works applies to the experimental case when
adequately reinterpreted \cite{Cuku}.
Let us also note that the in this paper the $\alpha$ and $\beta$ exponents
are exchanged with respect to Refs. \cite{Cukule}.
}
such as the  $p$-spin spherical spin glass \cite{Crhoso,Cukule}
or the model of a particle moving in an infinite-dimensional
random potential \cite{Kiho,Cukule},
$\alpha=1/2$ at $T=0$ and it decreases when increasing the temperature.
For the SK model, conversely, $\alpha = 1/2$ at $T=T_c$
and it decreases when decreasing the temperature \cite{Sozi}.
Finally, for the mixed $(p=2+4)$ spherical model introduced in Ref. \cite{Ni},
$\alpha$ has a non-monotonic dependence on $T$; $\alpha(T=0) = \alpha(T_c) =
1/2$.
The value of the exponent $\alpha$
measured  experimentally  follows the tendency of the one
holding for SK and the mixed $(p=2+4)$ spherical models close to the critical
temperature, though the value of $\alpha$ from the experiments is considerably
smaller
($\alpha \le 0.1$ {\it vs} $\alpha \sim 0.5$).

\subsubsection{Analytical Approach: Non-Stationary Regime.}

Following then very general requirements,
it has been argued in Ref. \cite{Cuku,Cuku3}, and explicitely checked on
several
pure and disordered models, that in the large-time limit only two situations
with
different dynamical behavior seem to exist:

\begin{itemize}

\item
On the one hand, there are models with only one time-scale -- or equivalently,
correlation-scale --
apart from the stationary one.
$C_A$ scales as in a domain
growth process within the non-stationary time-scale, in the sense that:
 \begin{equation}
C_A(t_w+t,t_w)
=
\jmath^{-1} \left( \frac{ h(t+t_w) }{ h(t_w) } \right)
\label{1blob}
\end{equation}
with $h(t)$ a monotonically increasing function -- analogous to the domain
length
$L(t)$. $\jmath^{-1}(u)$ is a function characterized by another exponent which
we call
here $(1-x)$ (and has been called $\beta$ in
\cite{Cukule} and $\alpha$ in  \cite{Bocukume}).  Close to $u =1$, {\it i.e.}
for the early epochs of the aging regime, $\jmath^{-1}$
reads
\begin{equation}
\jmath^{-1}(u) \sim q_{EA} - c (1-u)^{1-x}
\; ,
\label{earlyepochs}
\end{equation}
$c$ is a constant and $x < 1$ implying that $\jmath^{-1}$ is non-analytical in
the neighbourhood of $u = 1$
(see (\ref{pi1}) below).

In this case, the
FDT-violating factor $X[C_A]$ is a constant $X < 1$.
These models, when
treated statically with the replica trick, are solved by a
 {\it one step replica symmetry breaking} ansatz \cite{beyond}. An example is
the $p$-spin spherical model \cite{Crso}. We call them ``single-scale
models''.

Certainly the functions $h$ and $\jmath^{-1}$ do  depend on the specific model.
At the mean-field level we have succeeded in obtaining $\jmath^{-1}$ and $X$
for several models. However, surprisingly, there are for the moment no
analytical results available for the scaling function $h(t)$
\footnote{
This technical difficulty is related to the introduction
of a time re-parametrization invariance when studying the {\it exact}
mean-field
equations for large and widely separated times $t_w$ and $t+t_w$.
}.
The simplest possibility is that $h(t)$ is a pure power-law $(t/\tau_0)^a$, as
found in
the trap model \cite{jpbtrap} or standard coarsening models. This solution
is particular in the sense that  (\ref{1blob}) is then {\it independent} of the
microscopic time scale $\tau_0$, which can be taken to
zero; in this case $C_A$ simply depends on the ratio $t/t_w$ (full aging
situation, a qualitative approximation of the experimental results, as
explained in Sect. 2.1). This is not the case for more general functional
forms.  Two explicit choices which have been
proposed so far are:
 \begin{equation}
h(t) = \exp \left[\frac{1}{1-\mu} \left(\frac{t}{\tau_0} \right)^{1-\mu}
\right]
\qquad \mbox{or}
\qquad
h(t) = \exp \left[\ln^a(t/\tau_0) \right]
\; .
\label{mu}
\end{equation}
The form on the left was proposed to
account for
experiments in polymer glasses by Struik
\cite{struik}, then
used in the first accurate analyses of aging effects in the TRM-decay
\cite{aging2,nousaging}, and recently
found in the exact solution of the asymmetric spherical SK model with $\mu
=1/2$ \cite{Cukulepe}
(see also \cite{Frme}).
The second form is suggested by the numerical data from the ``toy model'' of a
point particle in a random
potential with infinite
dimension \cite{Cukule}. Both  will be used below to scale the data for the TRM
and the out-of-phase
susceptibility.
Note that in the limit
$\mu=1$ or $a=1$, one recovers full aging with a
pure power-law behavior for $h$, while $\mu=0$ corresponds to time translation
 invariance (no aging).
When $\mu<1$ and $a>1$ we have ``sub-aging'' that we define as follows.
Taking $t$ fixed and, say, in the beginning of the aging regime,
$h(t_w)/h(t+t_w) \sim 1- (d\ln(h(t_w))/dt_w) t$. This defines a
characteristic relaxation time $\tau(t_w)$. We say that we have sub-aging
(super-aging) when $\tau(t_w)$ grows slower (faster) than $t_w$.

Interestingly enough, one can in general derive a relation between $\alpha$,
$(1-x)$ and $X$ (see (\ref{FDTdecay}), (\ref{earlyepochs}) and
(\ref{X}) for their definitions)
\cite{Cukule,Bocukume}:
\begin{equation}
X \frac{ (\Gamma[1+(1-x)])^2 } {\Gamma[1+2(1-x)]}
=
 \frac{ (\Gamma[1-\alpha])^2 } {\Gamma[1-2\alpha]}
\; ,
\label{gammas}
\end{equation}
for single scale models. We shall use this equation to predict $X$
at the beginning of the beginning of the aging regime in Sect. 4.2.

\vspace{.5cm}

\item
On the other hand, there are models such as SK that have an infinite
number of time-scales -  correlation scales -
apart form the stationary one
\cite{So,Cuku3,Frme}; mathematically, one has
ultrametricity in time for all correlations such that $C_A < q_{EA}$, in the
sense that
$C_A(t_1,t_3) = \min( C_A(t_1,t_2),   C_A(t_2,t_3))$, $t_1>t_2>t_3$
\cite{Cuku3}.
The decay is here infinitely slower than in the single-scale models.
The FDT-violating factor $X[C_A]$ is a nontrivial function
of $C_A$.
These models are solved by a {\it full replica-symmetry breaking} ansatz
when using the replica trick at the static level \cite{beyond}, and can be
called ``multi-scale'' models.

\end{itemize}

It is important to notice that a scaling like (15) inside a correlation scale
and ultrametricity between different correlation scales are expected to
hold on very general grounds, in particular
for more realistic finite dimensional models (in the limit of large-times),
{\it provided that} the rather mild assumptions
used in \cite{Cuku,Cuku3} are satisfied.
This justifies
the fact that we shall use, in the following, a scaling-law like (\ref{1blob})
to scale the
data for real spin glasses, without refering to any particular model.

\subsubsection{Connection with Measurable Quantities.}

If linear response theory holds, the TRM is just
\begin{equation}
M(t+t_w,t_w) = h \int_{0}^{t_w} ds \, R(t+t_w,s)
\; .
\end{equation}
For large waiting-time $t_w$,
this integral can be rewritten using the decomposition
in stationary and non-stationary decays ((\ref{resp}) and (\ref{X}))
and using (\ref{FDTdecay}), for $t \gg \tau_0$ we have
\begin{equation}
\frac{M(t+t_w,t_w)}{M_{fc}(t+tw)} - A
\left( \frac{\tau_0}{t} \right)^\alpha
\propto
\int_{0}^{C_A} dC_A' \; X[C_A']
\; .
\end{equation}
where
$A$ is a constant.
For single scale models as in (\ref{1blob}) the scaling reads
\begin{equation}
\frac{M(t+t_w,t_w)}{M_{fc}(t+tw)} - A
\left( \frac{\tau_0}{t} \right)^\alpha
\propto
\jmath^{-1}
\left(
\frac{h(t_w)}{h(t+t_w)}
\right)
\; .
\label{TRMmeanfield}
\end{equation}
In the early epochs of the aging regime for the TRM, that should be compared to
the non-stationary
behavior of the $\chi''(\omega,t)$, one has
\begin{equation} \label{1-xM}
\frac{M(t+t_w,t_w)}{M_{fc}(t+tw)} - A
\left( \frac{\tau_0}{t} \right)^\alpha
\propto
q_{EA} - c \left(1- \frac{h(t_w)}{h(t+t_w)} \right)^{1-x}
\; ,
\end{equation}
where we used (\ref{earlyepochs}).

\vspace{.25cm}

The out-of-phase susceptibility can also be simply related to the
correlation function (\ref{autocorr}). For high-frequencies, $\omega t \to
\infty$,
the aging term does not contribute (it is
the integral of a slowly varying function $R_A(t,s)$ times
a rapidly oscillating function).  Using (\ref{FDTdecay}) to approximate the
remaining integral, one  finds the stationary part of the
a.c. susceptibility:
\begin{equation}
\chi''(\omega,t) \to  \chi_{eq}''(\omega) \propto
\omega^\alpha
\; ,
\;\;\;\;\;\;\;\;\;\;\; \omega t \to \infty
\; .
\end{equation}

Conversely, if $\omega t \ge 1$, {\it i.e.} for low frequencies,
the aging part strongly contributes. For single-scale
 models
we then have
\begin{equation}
\chi''(\omega,t)- \chi_{eq}''(\omega)  \sim
\frac{X}{T} h  \omega \int_{0}^t ds \exp(i\omega s) \; \jmath^{-1} \left(
\frac{h(s)}{h(t)} \right)
\; ,
\;\;\;\;\;\;\;\;\;\;\; \omega t \ge 1
\; .
\label{chisec}
\end{equation}
For $\omega t$ finite but large, one has in general:
\begin{eqnarray}
\chi''(\omega,t)- \chi_{eq}''(\omega)
\propto
  \jmath^{-1} (1) - \jmath^{-1} \left( 1- \frac{1}{\omega t} \, \frac{d \ln
h(t)}{d \ln t} \right)
\\
\label{corrX"}
\propto
\left( \frac{d\ln h(t)}{d \ln t} \, \frac{1}{\omega t} \right)^{1-x}
\;\;\;\;\;\;\;    1 \ll  \omega t < \infty
\; ,
\end{eqnarray}
where we introduced the power-law behavior of $\jmath^{-1}(u)$ in the vicinity
of $u=1$ defined
in (\ref{earlyepochs}). Hence the conclusions for $\chi"$:

\begin{itemize}
\item
If $h(t)$ is a simple power-law, then $\chi''(\omega,t)- \chi_{eq}''(\omega)$
scales as $\omega t$, as obtained in \cite{jpbtrap,jpbdd} (full aging).
\vspace{.25cm}

\item
If $h(t) = \exp(1/(1-\mu) t^{1-\mu})$ then  $\chi''(\omega,t)-
\chi_{eq}''(\omega)$ is a function
of $\omega t \times (t/\tau_o)^{\mu-1}$. When $\mu \neq 1$ there is
a correction to the pure $\omega t$ scaling (sub-aging for $\mu < 1$)).
\vspace{.25cm}

\item
If $h(t)$ is of the
form $\exp[\ln^a(t/\tau_0)]$, then the $\omega t$  scaling is corrected by a
slowly varying factor $\ln^{(a-1)}\left(t/\tau_0 \right)$ (sub-aging if $a >
1$). The $\chi"$ data are scaled within this assumption in the inset of Fig.2.
\end{itemize}

In Sect. 4, we apply the scaling relation (\ref{1blob}) from single-scale
models together with these proposals for $h(t)$
to the TRM and $\chi''$ data.

The mean-field models which lead to the above results
are very instructive: general statements and new ideas (like the violation of
FDT) have emerged from
their study. However, the physical mechanism underlying aging in these models
is not yet very clear.
The single-scale models, in particular, are very weakly sensitive to
temperature, suggesting that no
activated effects are involved, and that aging is rather related to large
dimensional
effects: the system wanders indefinitely in a large phase-space, without ever
reaching a local minimum
of the free energy \cite{Kula}.
Conversely, the trap model which we shall discuss now relies on activated
effects to generate a broad
distribution of time scales, which also leads to aging. However, this model is
phenomenological, and does
not emerge from a
precise microscopic description - although some steps in this direction have
recently been made \cite{jpbmm,leon}.
A tentative classification of the different models of aging has been proposed
in \cite{Babume}.

\subsection{The Trap Model}

In the trap model \cite{jpbtrap}, aging has been shown to
naturally occur in a situation called ``weak ergodicity breaking", which
corresponds here to a statistical
impossibility for the system to realize equilibrium occupation rates of the
metastable states. This model has
appeared as a fertile guideline for the analysis of the experiments; we
therefore recall its main points,
and come back to it later in Sect. 5.2. In the simplest version of the model
\cite{jpbtrap}, aging is sketched
by a random walk in a collection of ``traps" with random trapping times $\tau$,
all equally accessible. To each
trap is associated a certain magnetization M and ac susceptibility
$\chi_\tau(\omega)$. The properties of a real
sample are obtained by averaging over an ensemble of decorrelated {\em
subsystems}, corresponding to spins in
different regions of space
\footnote{
At this stage, the spins do not enter directly.
In later developments \cite{deltaH}, however, the number of spins to be flipped
for escaping from a trap
(and hence the size of the subsystems referred to above) has been estimated
from the influence of the field
amplitude on the dynamics, as observed in the experiments.
}.
 An important input is
common to microscopic theories \cite{beyond}, and also to the more general
problem of manifolds in random media
\cite{jpbmm,leon}; the distribution of trap depths is taken as an exponential.
For
thermally activated processes,
this yields the following distribution of trapping times:
\begin{equation}  \label{tau1+x}
\psi(\tau)={x\tau_0^x\over\tau^{1+x}}\ \ \ \ (\mbox{for}\ \tau\gg\tau_0)\ \  ,
\end{equation}
where $x$ (from the distribution of barrier heights) is a temperature dependent
parameter describing the
structure of the phase space. In a comparable way, the ``random energy model"
(REM) of Derrida \cite{REM}
involves $x=T/T_g$. The crucial point is that $x<1$ in the spin-glass phase; in
consequence, the mean value of
$\psi(\tau)$ is divergent, that is the mean time needed to explore the whole
set of traps (and thus to reach
ergodicity) is infinite. This was called weak ergodicity breaking, in the
sense that the equilibrium situation
is never realized, although the system never gets trapped in a finite region,
leading to an asymptotically zero correlation function -- see (\ref{Czero}).
This
 is very different of the usual ergodicity breaking, where the system can reach
rather quickly an equilibrium
configuration, but remains in a restricted sector of the phase space. Since
$x<1$, the distribution
(\ref{tau1+x}) is very broad. After a random walk during $t_w$, the system has
visited numerous short-life
traps, but in a relatively small time compared with $t_w$;
as usual with such broad distributions, the significant contributions
arise from the largest - although
rare - events. Thus, after $t_w$, the system has the largest probability to be
found in a trap of characteristic
time $t_w$ itself; if a magnetic field is varied at $t_w$, most subsystems will
need a time of order $t_w$ before
changing their magnetization. It has been shown  \cite{jpbtrap} that the
TRM-decay is then a function of $t/t_w$,
and similarly that the ac susceptibility is a function of $\omega.t$.
Thus, on the basis of a statistical description in the space of the metastable
states, the main features of
aging can be obtained.  Further developments of this approach
\cite{jpbdd,jpbev} are discussed below at
the light of various aspects of the experimental results. We now turn to a more
detailed description of the
combination of aging {\em and} stationary dynamics as seen in {\em both} TRM
and $\chi"$  measurements.

\vspace{2cm}
\section{TRM Experiments: Aging and Non-Aging Dynamics Disentangled}

\subsection{Departures from Full $t/t_w$ Aging}

In Fig.1.b, the TRM curves are presented as a function of $t/t_w$; it is clear
that this is not exactly the correct reduced variable
(failure of a  full aging scaling). The same
effect has been found in other samples \cite{aging2,nousaging} and can be seen
in
results from other laboratories
({\it e.g.} \cite{lundgren1983}). It has been first identified in polymer
mechanics \cite{struik}. Indeed,
the $\chi"$ results recall us that stationary dynamics, namely
$\chi"_{eq}(\omega)\propto\omega^\alpha$
(an additive contribution to the aging part, see Fig. 2), must intervene in the
TRM decay, particularly in the
$t\ll t_w$ regime. This frequency-dependent  $\chi"_{eq}(\omega)$ is equivalent
to an additive
contribution of the form $t^{-\alpha}$ to the TRM.

The question of the departure from full aging
has already been discussed in the past and the TRM's have been
very accurately parametrized
as the {\em product} of a $t^{-\alpha}$ factor times a $(t,t_w)$ dependent
factor
(\cite{nousaging}, see also the
footnote$^2$ above). However, an additive combination of the stationary and
aging parts arises naturally
in the above theoretical approaches \cite{Cuku,jpbtrap}, and we shall reanalyze
these
data in this light. We express the stationary part of the TRM, in units of the
field-cooled
value $M_{fc}$ like in Fig. 1, as $A(\tau_0/ t)^\alpha$. $\tau_0$ is again a
microscopic time, which allows
homogeneity of the units, and then A is a non-dimensional constant, expected to
be of order 1. We have taken
the same TRM-data as in Fig. 1, and adjusted A and $\alpha$ in order to try
to merge
the aging parts $f(t_w,t)$
\begin{equation} \label{A+B}
{M\over M_{fc}}=A \left({\tau_0\over t} \right)^\alpha+f \left(\frac{t}{t_w}
\right)
\end{equation}
 of all 5 curves of various $t_w$'s as a function of $t/t_w$. 
We have no $\alpha$ values from $\chi"$-measurements on the $AgMn$ sample, so
we have simply kept $\alpha$ in the
$0.01-0.1$ range obtained for the $CdCr_{1.7}In_{0.3}S_4$
sample in previous analyses \cite{nousaging}. Fig. 3.a shows the result;
whatever the choice of parameters (also,
trying a logarithmic decrease instead of a power law), it seems impossible to
merge all 5 curves together
in the {\em whole time regime}. Systematic discrepancies are always found in
the
large $t/t_w$ range: no simple function of $t$ alone can account for these
deviations. However, for small $t/t_w$, the
good quality of the scaling {\it is equivalent} to the simple
$\omega.t$-dependence
of  $\chi"(\omega,t)$, which is actually
measured in the {\em same time regime} ({\it i.e.} $\omega t>1$).


\setcounter{figure}{2}
\begin{figure}
\centerline{\epsfxsize=10cm
\epsffile{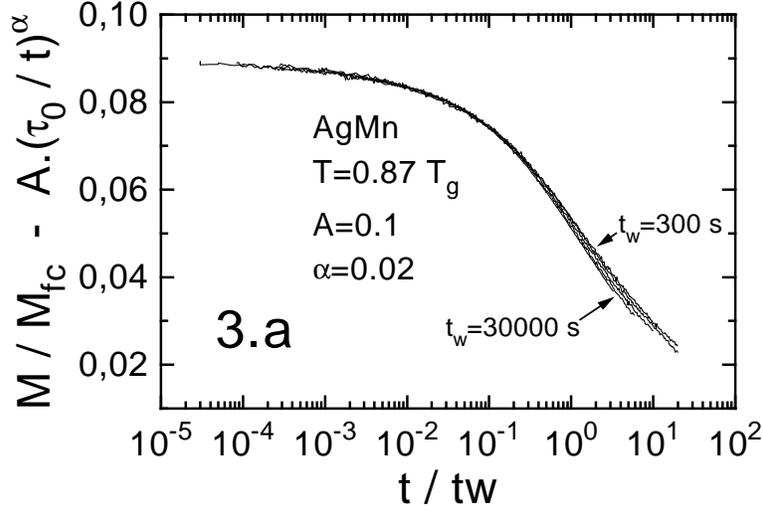}
}
\vspace{.2cm}
\caption{{\bf a.}
Aging part of the TRM (Eq.\ref{A+B}): the estimated stationary contribution has
been subtracted from the full measured value. The data (same as in Fig. 1ab) is
plotted {\it
vs.}~$t/t_w$.
}
\end{figure}


In the long time $t/t_w>1$ region, the discrepancies shown in Fig. 3.a present
the same features - although somewhat weaker - as in
previous analyses using a multiplicative stationary contribution
\cite{nousaging}.
In Fig. 3.a, as well as in Fig. 1.b (the same, without stationary dynamics
subtraction), it is clear that a
$t/t_w$-scaling slightly over\-estimates the aging effect
(sub-aging); the ``youngest" curve (shortest $t_w$) lies
{\em above} the others, whereas it was {\em below} in the raw data (Fig. 1.a),
and the effect is sytematic for
the 5 curves.

\subsection{Sub-Aging Scaling}

A way to succeed in merging all TRM curves is to use a generalized scaling
function of the
form $h(t)=\exp[\frac{1}{1-\mu}(\frac{t}{\tau_0})^{1-\mu}]$
with $\mu < 1$
(one of both examples quoted above in (\ref{mu})). This was proposed in the
context of spin glasses in
\cite{aging2,nousaging}, as a phenomenological scaling procedure inspired from
polymer
mechanics \cite{struik}, which proved
to account with great accuracy for the observed sub-aging situation. We refer
the reader to \cite{aging2,nousaging} for the
arguments which lead to the above  choice of $h(t)$, or rather, along the lines
of \cite{nousaging}, to the {\em effective time}
$\lambda$ defined as
\footnote{
The careful reader will notice a change of $\lambda$-units when compared with
\cite{nousaging}; 
within the present definition, $\lambda$  is equal to
$\lambda(\frac{\tau_0}{t_w})^{\mu}$ from
\cite{nousaging}.
} :
\begin{equation} \label{lambda=}
\frac{h(t+t_w)}{h(t_w)}=\exp{\frac{\lambda}{\tau_0}} \quad .
\end{equation}
As a function of $\lambda$, the 5 curves in Fig. 3.a recover precisely the same
shape \cite{aging2,nousaging},
as displayed in Fig. 3.b. In other words, $\lambda$ results from a change of
variable
which allows us to see the aging dynamics
as {\em stationary}. It can also be obtained from
\begin{equation}
\frac {d\lambda}{\tau_0^\mu}=\frac{dt}{(t_w+t)^\mu} \quad ,
\end{equation}
which means that, since the age $t_w+t$ is varying during the TRM experiment,
the effect of an elementary time interval
$dt$ depends on the elapsed time;   
 due to sub-aging, the effective time scale associated to the age $(t_w+t)$ is
$(t_w+t)^\mu$, with $\mu <1$.


\setcounter{figure}{2}
\begin{figure}
\centerline{\epsfxsize=10cm
\epsffile{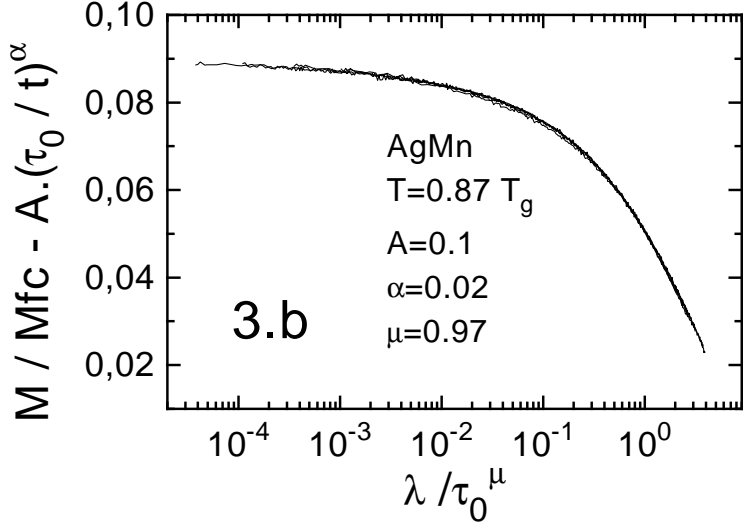}
}
\vspace{.2cm}
\caption{{\bf b.}
Aging part of the TRM as in Fig. 3a, but {\em vs} the scaling variable
$\lambda/\tau_o^\mu$ defined in  (\ref{lambda=}) (1st example in (\ref{mu})). 5
curves, obtained for $t_w=300,1000,3000,10000,30000 s$ (same data as in Fig. 1
and 3a), are superimposed onto each other.
}
\end{figure}


This type of scaling has been successfully  applied to various samples
\cite{aging2,nousaging}; $\mu$ is found in the $0.8-0.9$ range and it is
almost independent of temperature in the $0.3<T/T_g<0.9$
range. This $\mu$-trick is a very convenient way of parametrizing the
sub-aging deviations from a full $t/t_w$-scaling,
although not necessarily having a direct physical meaning. $\mu=0$ corresponds
to the case of no $t_w$-dependence (no aging),
while $\mu=1$ would yield a full $t/t_w$ scaling. The intermediate values of
$\mu$ (sub-aging)  correspond to the fact that
the apparent relaxation time scales sub-linearly (as $(t_w+t)^\mu$) with the
age $t_w+t$.

The alternative choice of $h(t)$ proposed above in (\ref{mu}) can also be
considered
\cite{Cuku4}. It is actually very close to the $\mu$-case, in particular in
the limit $\mu \to 1$. Again, for $a=1$, full aging is recovered, while $a>1$
describes a sub-aging situation, with an apparent relaxation time $\tau(t_w)=$
$t_w/(a \ln^{(a-1)}(t/\tau_0))$. We have also applied this other
sub-aging scaling to the non-stationary part of the relaxation displayed in
Fig. 3a; the result is
shown in Fig. 3.c, again the curves are fairly well superimposed onto each
other.

Eq.(\ref{gammas}),
derived for single-scale mean-field spin-glass models,
relates $\alpha$, $x$ and the FDT violating factor $X$ defined in (\ref{X}).
According to (\ref{1blob}) and (\ref{earlyepochs}), we have made a fit of the
beginning of the TRM decay in Fig. 3.c  (early epochs), and we have obtained $x
\sim 0.96$ (see (\ref{earlyepochs})). This is consistent with $X=1$, which
would suggest that FDT applies without corrections even at the beginning of the
aging regime, if one accepts that
(\ref{gammas}) holds. This would be in accord with previous indications
obtained
through the comparison of $\chi''$ and noise measurements  made in the
early epochs of the aging regime (``quasi-stationary regime") \cite{nousaging}.
A careful and detailed study of the noise auto-correlation and response
functions would help us knowing if and how
FDT is violated, namely if $X$ departs from 1 when going deeper in the aging
regime. .


\setcounter{figure}{2}
\begin{figure}
\centerline{\epsfxsize=10cm
\epsffile{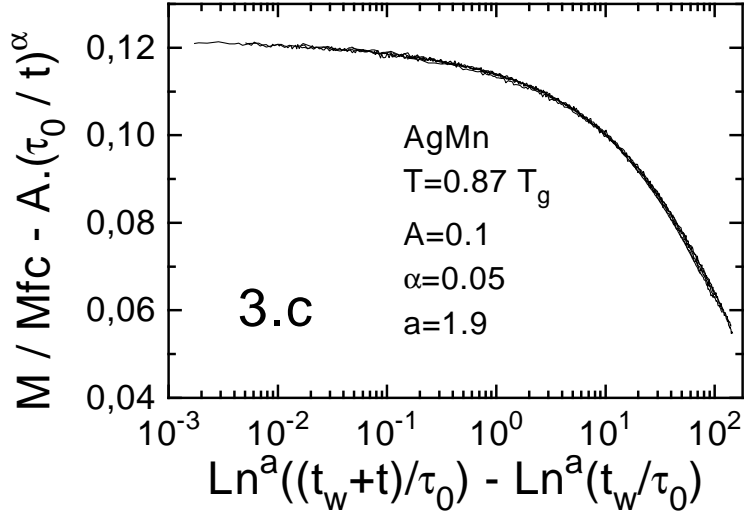}
}
\vspace{.2cm}
\caption{{\bf c.}
Same as Fig. 3b, but as a function of a scaling variable which corresponds to
the 2nd example in (\ref{mu}).
}
\end{figure}


For the sake of comparison,
we show in Fig. 3d how the $\mu$-scaling applies to the full value of the
TRM (the stationary
contribution is neglected, and {\em not subtracted} from the magnetization). An
almost acceptable scaling
is obtained; the discrepancies are not much larger than the experimental
uncertainties, but {\em they are systematic}:
all curves are crossing in the middle of the figure. This corresponds to
$\mu=0.87$. Neglecting the influence of the
stationary contribution thus enhances the deviations from $\mu=1$.


\setcounter{figure}{2}
\begin{figure}
\centerline{\epsfxsize=10cm
\epsffile{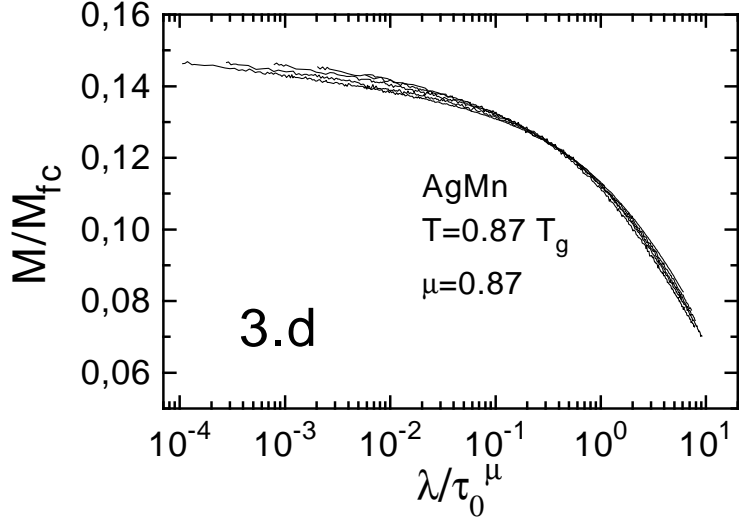}
}
\caption{{\bf d.}
Full measured value of the TRM (as in Fig. 1ab), but {\em vs} the scaling
variable $\lambda/\tau_o^\mu$ defined in  (\ref{lambda=}).
}
\end{figure}


The very good scaling of the 5 curves in Figs. 3.b, 3.c,
involves 3 free parameters: here $A=0.1,\alpha=0.02,\mu=0.97$
or $A=0.1,\alpha=0.02,a=2.2$.
They  correspond to a full account of the results, with good
coherence between TRM and $\chi''$ experiments. Let us summarize our points:

\begin{itemize}

\item
The out-of-phase susceptibility
$\chi"(\omega, t)$ is
well represented by the sum of a stationary contribution, which varies like
$\omega^\alpha$ (or $(\ln \omega)^{power}$), plus a
non-stationary contribution which varies as a negative power $x-1$ of
$\omega.t$
(possibly with slowly varying corrections
suggested by (\ref{corrX"}) above, but which are hardly visible on the $\chi''$
data; see the rescaled $\chi"$ in the inset of Fig.2).

\vspace{.5cm}

\item
The TRM decay curves are  the sum of a
stationary contribution $\propto  t^{-\alpha}$, plus an aging contribution. In
the time range ($t\ll t_w$) which
is common to $\chi"$-measurements, this aging part can be approximated (in
agreement with $\chi"$) by a
$t/t_w$-scaling. The long time regime $t\ge t_w$ seems however to prefer a
sub-aging scaling of
the type given in (\ref{mu}).

\end{itemize}

The origin of this weak ($1-\mu=0.03$) but insistant sub-aging behavior is an
interesting point, still
uncompletely understood. As mentionned above, from a theoretical point of view
this would mean that
the $\tau_0 \to 0$ limit of the underlying model does not exist. From a more
physical point of view, several
scenarios which might explain this effect can
be considered. The simplest one concerns the effect of the field amplitude,
which indeed (for larger fields)
is known to suppress progressively the $t_w$-effect {\footnote{In a similar
way, $\mu$ is seen to decrease
with the stress amplitude in polymer glasses \cite{struik}}}(see \cite{deltaH}
for details). However, the
parameter $\mu$ seems to stick to a plateau value ($<1$) for
 the explored low-field range of $10-0.1 Oe$ (systematic scaling analysis with
{\em additive} (rather than
multiplicative) stationary corrections are however needed to confirm this
point).

This sub-aging behavior might also be interpreted as a sign that aging is
actually ``interrupted" beyond
very long, but finite, times \cite{jpbev}. For
example, if the trapping time distribution is cut-off beyond a certain ergodic
time $t_{erg}$ (which itself
depends on the subsystem), then {\em for part of the subsystems} ergodicity
will be realized within the
time of the measurement: their dynamics will no more depend on $t_w$
(interrupted aging). As shown
in \cite{jpbev},
this produces an effective sub-aging scaling very close to the $\mu<1$ effect
described here. The result is a value
of a typical
time $t_{erg}$, which must be understood as a {\em crossover time scale},
beyond which $\mu$ will further
decrease to $0$.
In the analysis of \cite{jpbev}, however, we had not properly taken into
account the stationary contribution,
which we have shown here to bring
$\mu$ much closer to 1 (Figs. 3.b and 3.d). We had found $t_{erg}$ of the order
of $10^{6-7}sec$, which might
therefore be underestimated.

\vspace{2cm}
\section{Towards a Hierarchical Description of the Space of the Metastable
States}

\subsection{Temperature Variation Experiments}

\subsubsection{Main Qualitative Features: ac Measurements.}

Until now, we have only presented results which are obtained from aging
experiments {\em
at  constant temperature}
after the quench into the spin-glass phase. In another class of experiments,
the temperature is varied
during aging. In terms of thermally activated processes in a mountaneous
free-energy landscape, one may expect to explore in more detail the various
scales of
the free-energy barriers which
are involved in the slow dynamics. Thermal activation should - at first sight -
be
able  to speed up or slow down
the aging evolution. The conclusions of these experiments have been
surprisingly instructive.

An early result was obtained in \cite{lundXdT}; measuring the aging
relaxation of $\chi"$ in a CuMn
spin glass, the authors observed that any step increase or decrease in
temperature was causing an
instantaneous increase of $\chi"$, followed by a slow decrease. We may call
this effect ``restart of aging",
since the relaxation is renewed by the temperature change, which hereby
produces a similar effect as
obtained after the quench. This phenomenon reveals that the slow aging
evolution towards equilibrium
is significantly disturbed by relatively small temperature changes. Such a
``chaotic dependence" of the
equilibrium states on temperature has  been predicted to occur as a consequence
of frustration in
\cite{fhbm}, where it is argued that the relatively small free-energy of an
overturned region of spins
(droplet) results from large cancellations at the surface of the droplet.
These cancellations should be
very sensitive to temperature, hence the strong effect of a small temperature
variation.

When studied in more detail,
the temperature variation experiments do not simply show a  restart of aging
upon {\em any} temperature
change. We present in Fig. 4 an experiment performed in such a way that the
reaction of the $\chi"$-relaxation
to either a {\em decrease} or an {\em increase} in temperature is very
different \cite{jphys+}. The sample is first
quenched from above $T_g=16.7\ K$ to 12 K; due to aging, $\chi"$ slowly
relaxes. After 350 min, the temperature
 is decreased to 10 K. Despite the reduced thermal energy, the relaxation does
not slow down, but restarts
abruptly from a higher value, in agreement with \cite{lundXdT}; this is the
surprising chaotic-like effect, which
looks as if the system was (at least partially) restarting aging from the
quench.
But, when after another 350 min at 10 K the sample is heated back to 12 K, the
result is
very different: $\chi"$ resumes its slow relaxation from the value which had
been reached
before the temperature variation (see the inset of Fig. 4), in a {\em
memory-like} effect.


\setcounter{figure}{3}
\begin{figure}
\centerline{\epsfxsize=11cm
\epsffile{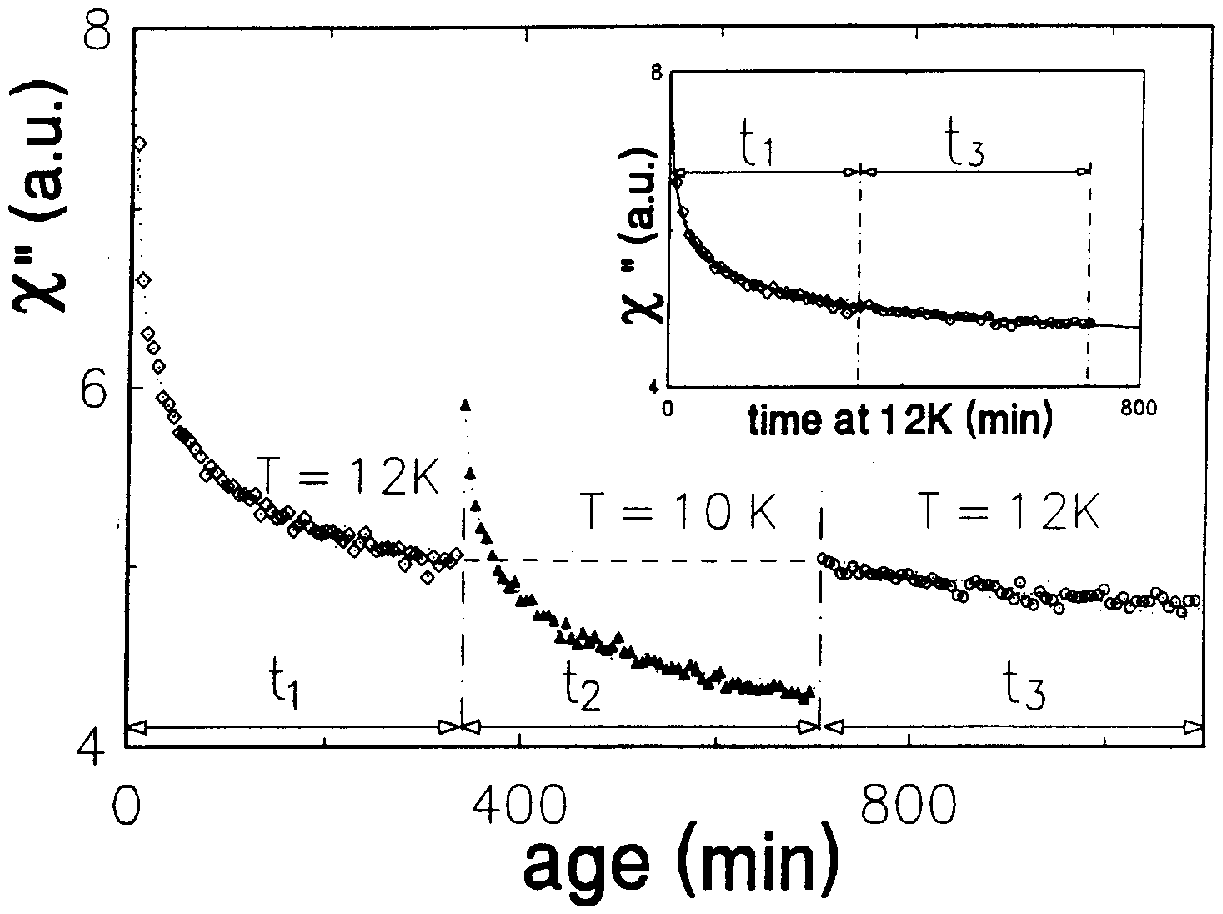}
}
\caption{
Out of phase susceptibility $\chi''(\omega,t_a)$ of the
$CdCr_{1.7}In_{0.3}S_4$ sample ($T_g=16.7K$) during a  temperature cycle. The
frequency $\omega$ is 0.01 Hz, and $t_a$ is the time elapsed from the quench.
The inset shows that, despite the strong relaxation at 10 K, both parts at 12 K
are in continuation of each other.
}
\end{figure}


There is no contradiction with the results in \cite{lundXdT}. The experimental
conditions of
Fig. 4 are chosen here as the most illustrative of this twofold effect. Indeed,
for smaller
temperature variations, more intricate situations can be found \cite{andalo};
aging at the
lower temperature may partly contribute to aging at the higher temperature,
even with a
temporary incoherent transient, as emphasized {\it e.g.} in \cite{uppsaladT}.
What we want to stress here is that a restart of aging is always observed after
a temperature decrease;
in contrast, what is found upon a temperature increase is
a memory of previous aging {\em at this higher temperature}. On the one hand,
if a long time has
been spent previously at
the higher temperature, as is the case in Fig. 4, only a weak relaxation is
found; on the other hand, if the
system has been {\em directly} quenched to a given temperature, heating up
afterwards to a higher
temperature - for which no memory of previous aging exists - will produce a
strong relaxation at this
temperature (see examples of various situations in \cite{andalo}).

\subsubsection{Same Effects in TRM Experiments.}

The effect is quantitatively confirmed by measurements of the TRM-relaxation
\cite{ryan,jphys+}.
The comparison with $\chi"$ requires some care; the $\chi"$-relaxation directly
shows {\em on-line} aging
as a function of time, whereas in the TRM the effect of aging during $t_w$ is
considered {\em afterwards},
during the relaxation which follows the field cut-off at $t_w$. The TRM curve
shows the relaxation
processes in a wide time window, and thus yields more extensive information
than $\chi"$ at a given
frequency $\omega$ (which mainly reflects the processes of characteristic time
$\simeq1/\omega$).
 In Fig. 5.a, a negative temperature cycling has been applied during the
waiting time. 


\setcounter{figure}{4}
\begin{figure}
\centerline{\epsfxsize=11cm
\epsffile{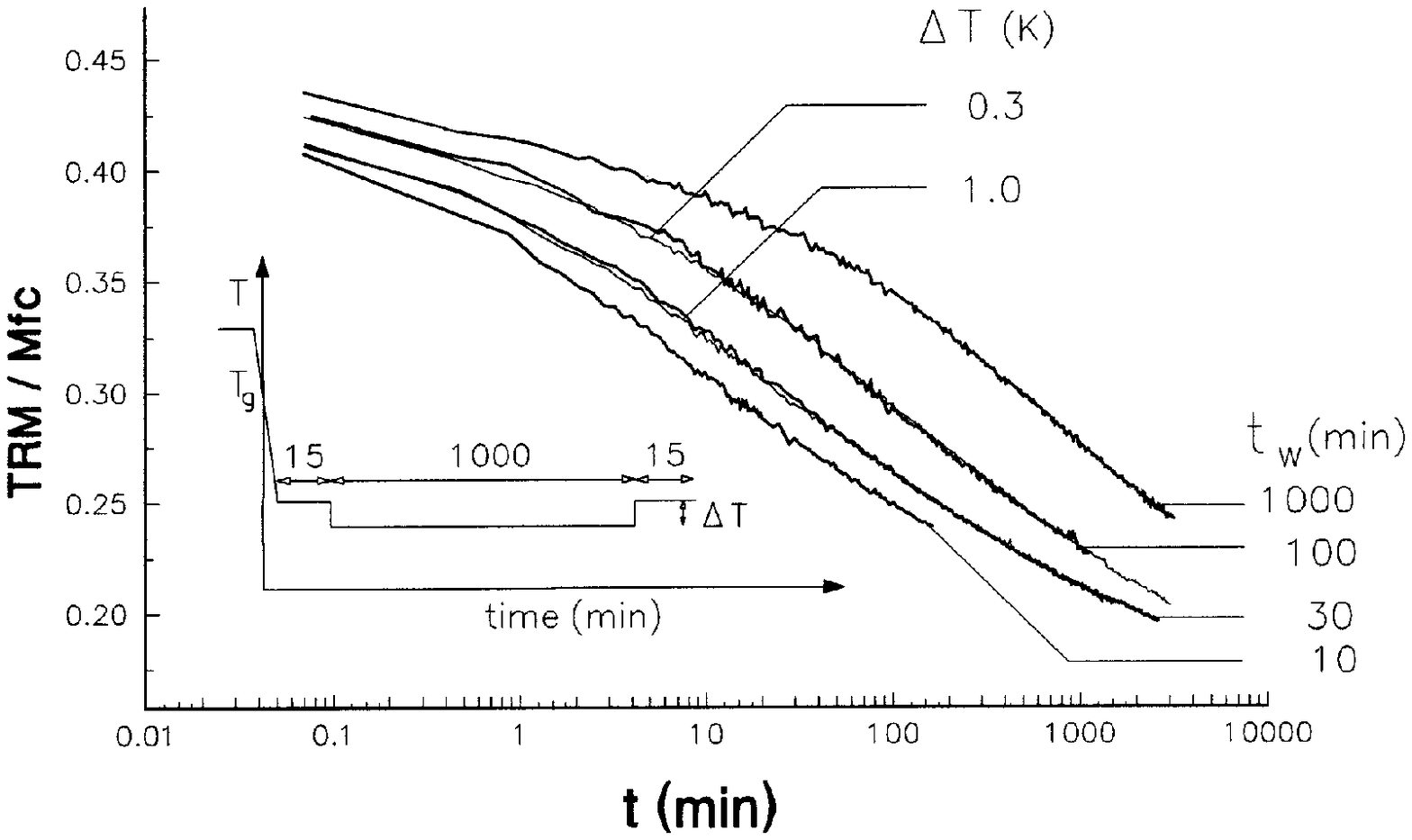}
}
\caption{{\bf a.}
Effect on the TRM relaxation of a {\em negative} temperature cycle
($CdCr_{1.7}In_{0.3}S_4$ sample, $T_g=16.7 K$). After waiting $t_{w+}=15 min$
at $T_0$ ($=12K=0.7T_g$), the sample is cooled to $T_0-\Delta T$
for $t_w-=1000 min$, and then is
heated back to $T_0$; after another $t_{w+}=15 min$, the field is cut and  the
relaxation measured (at $T_0$).
These relaxations  (thin lines) are compared with normal ones, measured after
waiting
$t_w=10,30,100 or 1000 min$ at
constant $T_0$ (bold lines).
}
\end{figure}


{}From the above $\chi"$ results, one expects that aging
processes be restarted
when going to $T_0-\Delta T$; but for sufficiently large $\Delta T$ this
evolution will be erased when
coming back to $T_0$. This is what can be checked in Fig. 5.a for $\Delta
T=1K$: the procedure yields
a relaxation curve which is exactly superimposed onto that obtained after
simply
waiting $t_{w+}+t_{w+}=30 min$ at
constant $T_0$, aging during $t_{w-}=1000 min$ at 11 K has not contributed.
Note that
the equivalent normal
curve has $t_w=30 min=2t_{w+}$, not $15 min$ (see Fig. 5a); the memory of the
first
aging stage has indeed been preserved.

Intermediate $\Delta T$ values produce intermediate situations; for $\Delta
T=0.3K$, the resulting curve is
the same as that obtained after waiting $t_{eff}=100 min$ at constant $T_0$.
{}From this
example, we can work out a
quantitative discussion of the effect. 
If we consider that the same characteristic free-energy barrier has been
crossed by thermal activation during

i) $t_{w+}+t_{eff}+t_{w+}=100min$ at $T_0$ and

ii) ($t_{w+}$ at $T_0$) + ($t_{w-}$ at $T_0-\Delta T$) + ($t_{w+}$ at $T_0$),
\\ \noindent
then we can write

\begin{equation} \label{ralent}
(T_0-\Delta T)\ln {{t_{w-}}\over {\tau_0}} = T_0 \ln {t_{eff}\over\tau_0} \quad
,
\end{equation}
which yields for the attempt time $\tau_0$ the unpleasant value of $\sim
10^{-42} s$.
Obviously, the {\em memory effect} cannot be explained by the only thermal
slowing down of jumping processes over constant height barriers.

In addition, simple thermal effects cannot be expected to explain the restart
of
aging, which is again evidenced in 
the positive cycling procedure of Fig. 5.b (all temperatures remain below
$T_g$).


\setcounter{figure}{4}
\begin{figure}
\centerline{\epsfxsize=11cm
\epsffile{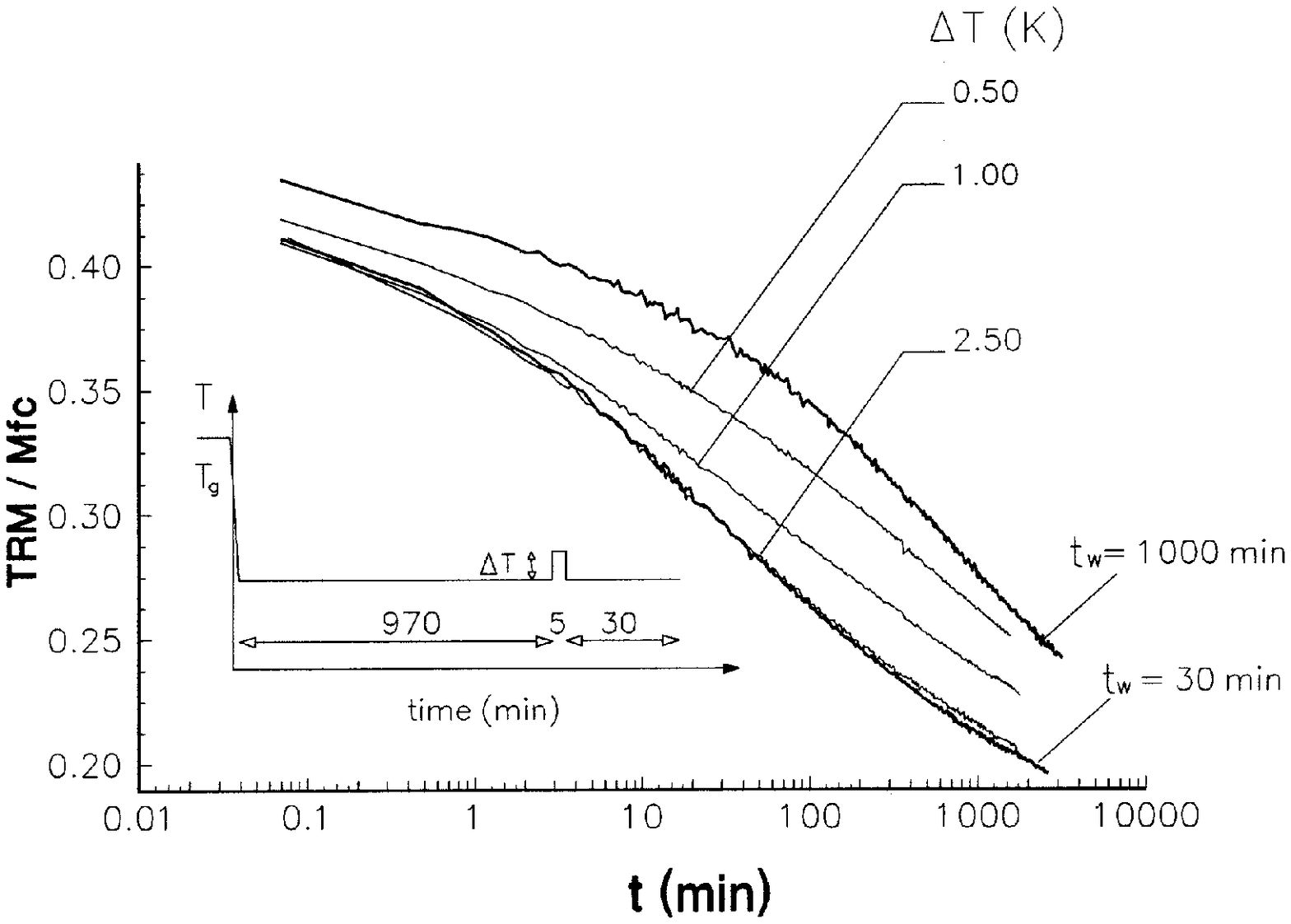}
}
\vspace{.2cm}
\caption{{\bf b.}
Effect on the TRM relaxation of a {\em positive} temperature cycle
($CdCr_{1.7}In_{0.3}S_4$ sample, $T_g=16.7 K$). A short heating cycle is
applied after 970 min of waiting time at $T_0$; then one still waits 30 min
before cutting the field and
measuring the relaxation at $T_0$ (thin lines). These relaxations are compared
with normal ones, measured after waiting
$t_w=30\  {\mbox or}\ 1000 min$ at
constant $T_0$ (bold lines).
}
\end{figure}


 What appears in this procedure is the
restart of aging due to the
temperature decrease at the end of the cycle; for a sufficient $\Delta T$
($=2.5 K$), the restart  is so
strong that the 970 min of previous aging are completely erased, as proved by
the superposition of the
$\Delta T=2.5K$ curve with a normal $t_w=30 min$ one. 

Again, for intermediate
$\Delta T$ values, the
effect is only partial; in its short-time part, the $\Delta T=1K$ curve sticks
to the young $t_w=30 min$-curve,
whereas in its long-time part it goes closer to older ones. In contrast with
the result in Fig. 5.a (negative
T-cycling), the present procedure (positive T-cycling) yields curves which are
not equivalent to a given
waiting time at constant temperature. Clearly, the reason for that is that the
positive temperature cycle
ends by cooling down to $T_0$ (which produces a partial restart of aging at
$T_0$), whereas the negative
temperature cycle ends by heating back to $T_0$ (which lets the system retrieve
the memory of the previous
aging at $T_0$).

Thus, the same features of aging are observed in $\chi"$ and TRM experiments
\cite{jphys+,ryan}.
A chaotic nature \cite{fhbm} of the spin-glass phase appears when the
temperature is
decreased, the restart of aging processes being very similar to
what initially happens after the quench. On the other hand,  a memory effect
is found when the
temperature is raised back, and this memory effect goes far beyond what can be
expected from thermal slowing
down. 

\subsubsection{A Hierarchical sketch.}

These effects have been interpreted  in terms of a {\em hierarchical
organization} of the
metastable states as a function of temperature \cite{ryan,jphys+}.
The empirical picture is sketched in Fig. 6.


\setcounter{figure}{5}
\begin{figure}
\centerline{\epsfxsize=10cm
\epsffile{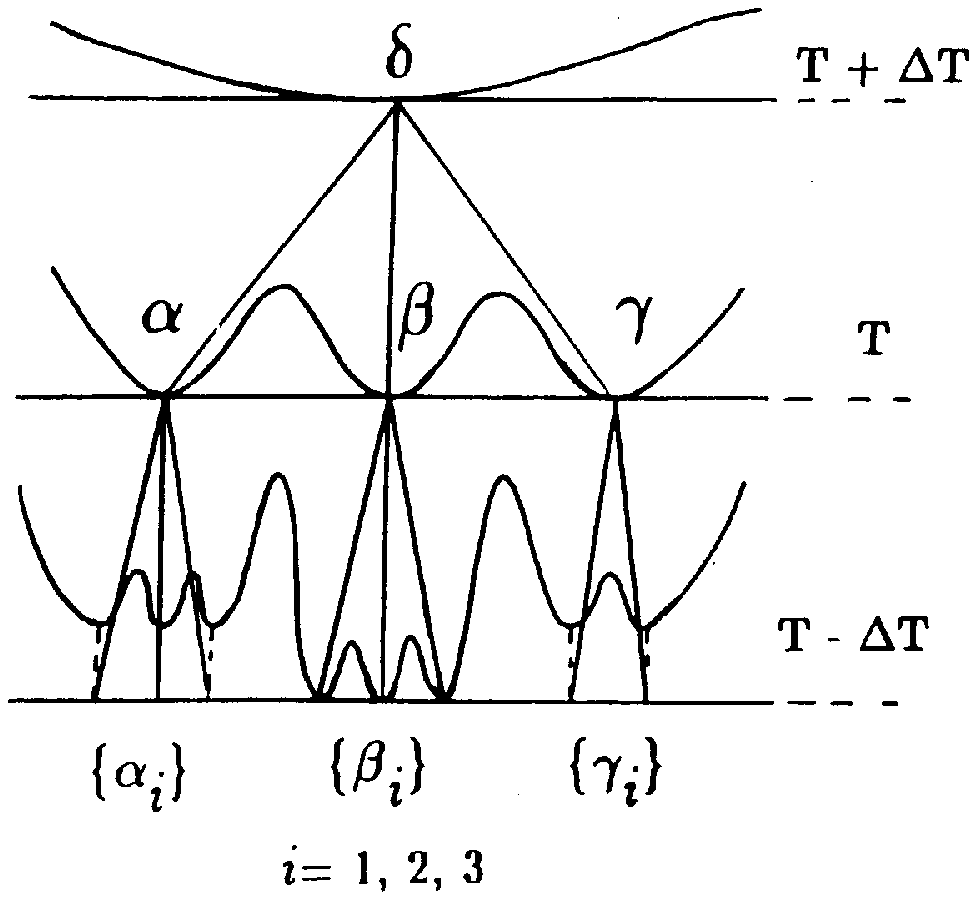}
}
\caption{
Schematic picture of the hierarchical structure of the metastable states as a
function of temperature.
}
\end{figure}


During aging at a given T,
the system samples the free-energy valleys (metastables states) at a given
level of a hierarchical tree;
the restart of aging upon lowering the temperature is figured out as a
subdvision of the free-energy
valleys into others, pictured as the nodes of a lower level of the tree, which
thus develops multi\-furcating
branches as the temperature decreases. The system is partially quenched
since it must now search for
equilibrium in a new, unexplored landscape, therefore aging (at least
partially) restarts. The hierarchical picture
naturally provides us with the observed memory effect; when the temperature is
raised back, the newly born
valleys and barriers merge back to the previous free-energy landscape. Thus,
aging at $T-\Delta T$ may have
{\em not contributed}  to the evolution at T, as far as all sub\-valleys which
could be explored at $T-\Delta T$
originate from unique valleys of the landscape at T  (which is
realized for large enough $\Delta T$).

The interpretation of these effects in a picture where aging is seen as the
growth of {\it compact}, independent
 domains (droplets \cite{FH}, domains \cite{KH}) remains, in our
opinion, difficult
\footnote{
See \cite{ryan,jphys+}. However, some experiments have been analyzed along this
line, with the introduction of long-time effects for the breakup of the domains
\cite{KH,uppsaladT}.}
{}.
 The droplets may be broken into smaller ones when the temperature is
decreased, thus producing a restart of
aging; but the memory effect implies that some information is kept somewhere
about the {\em stage of aging
which had been reached before decreasing the temperature}.
Thus, large-scale correlations should be kept untouched, while on the other
hand they should apparently be
destroyed. A way to satisfy these requirements could be  to
consider that the droplets should
be fractal (non space filling) \cite{clusterocio,villain}, with a non trivial
internal structure, in such
a way that large droplets may contain
smaller ones which can be activated independently. These
droplets inside droplets might lead, under some energetic conditions, to a
space-transcription of the hierarchical organization of the states.

\subsubsection{More Quantitatively: Rapid Growth of Free-Energy Barriers.}

The observed effect on aging of various thermal histories is in contradiction
with thermal activation over {\em constant height} barriers, as evidenced above
in (\ref{ralent}). The picture can be made more quantitative along this same
line
(Fig. 5.a and \ref{ralent}), which we now further develop. In one experiment,
the spin
glass is aged during $t_w^-$ at $T-\Delta T$, and brought back to T before
cutting the field and measuring the
relaxation. In another experiment, the spin glass is simply aged during $t_w^0$
at T, and the relaxation is measured
at T. If both decay curves are the same in the whole time window, we may
consider that the aging states reached in both procedures are the same, and
that equivalent
regions of both landscapes at
$T-\Delta T$ and T have been explored. We can characterize each evolution by a
typical
height B of the maximum barrier
which has been crossed in each case, namely
\begin{eqnarray} \label{barrier}
B(T-\Delta T)=(T-\Delta T).\ln{t_w^-\over\tau_0} \\
B(T)=T.\ln{t_w^0\over\tau_0}\ \ .
\end{eqnarray}
Thanks to the identity of the TRM curves obtained in each procedure, one may
consider that this  quantifies the T-variation of the {\em same barrier}, which
limits in both
cases the {\em same region} of the phase space. The contradiction pointed out
in (\ref{ralent}) shows that
$B(T-\Delta T)>B(T)$; in other words, the hierarchical
picture of valleys bifurcating into valleys  as the temperature decreases is
supported by the observation of the growth of free-energy barriers.

An extensive series of TRM measurements on the AgMn sample,
for multiple values
of $t_w^-$, $T$ and $\Delta T$,
has been performed \cite{ucladT}. In brief,
the result is that the barriers are
growing for decreasing temperatures below $T_g$, and in addition that their
growth rate is so fast that {\em at any $T<T_g$} some of them should even {\em
diverge} (see details in \cite{ucladT}).  This provides us with an interesting
link to the Parisi solution of the mean-field spin glass \cite{beyond}.
For decreasing
temperatures, some barriers separating the metastable states are diverging,
transforming the valleys into pure states in the sense of the Parisi solution;
the hierarchical structure of the
valleys, deduced from the experiments, can thus be related to
 that of the pure states. Also, the picture which emerges from these results is
that of
a {\em critical regime} at any $T<T_g$; in a sequence of micro-phase
transitions starting at $T_g$, the spin-glass phase space continuously splits
into nested and mutually inaccessible regions, within which non trivial
dynamics takes place.

\subsection{Aging as a Random Walk: Traps on a Tree}

\subsubsection{From a One-Level to a Multi-Level Tree.}

In its first stage \cite{jpbtrap}, the trap model deals with all-connected
traps, which can be called a ``one-level  tree" (see Sect. 3.3).
 The basic quantity which is calculated is the spin-spin correlation function
$C_A(t_w+t,t_w)$, which can be
related to the relaxation function if the fluctuation-dissipation theorem
holds  
(in non-equilibrium, generalized
forms of FDT might nevertheless hold \cite{Cuku}, see Sect. 3.2). The aging
part $M_A$ of the TRM-relaxation can thus be estimated from the
decay of the aging correlation function, which is proportional (by a factor
$q_{EA}$)
to the probability $\Pi(t,t_w)$ that the system has not jumped out of a
$t_w$-trap at $t_w+t$:
\begin{equation} \label{Ctt}
M_A(t+t_w,t_w)\sim C_A(t_w+t,t_w)=q_{EA}.\Pi(t,t_w)\ \ .
\end{equation}
The probability $\Pi$ contains all the information on possible jumps from trap
to trap, and thus represents the aging dynamics; it should tend to 1 when $t_w$
goes to infinity, for any finite $t$. In this limit, equilibrium dynamics is
recovered, since no jump occurs; the proportionality factor $q_{EA}$ in
(\ref{Ctt})
describes this ``bottom of the traps" dynamics ({\it cf.} with
(\ref{FDTdecay}) in Sect.3.2). From its definition, it
represents the overlap between the various configurations which constitute the
bottom of the traps; in a TRM experiment, it is this
finite fraction of the initial magnetization which should be found in the ideal
limit of infinite $t_w$.

Two asymptotic behaviors of $\Pi(t,t_w)$ (and thus of the TRM) are calculated
in a ``multi-level" version of the trap model \cite{jpbdd}:
\begin{eqnarray} \label{pi1}
t\ll t_w  \qquad \Pi(t,t_w) &\sim& A-\left({t\over t+t_w}\right)^{1-x_M} \\
t\gg t_w \qquad \Pi(t,t_w) &\sim& \left({t\over t+t_w} \right)^{x_1}\ \
,\nonumber
\end{eqnarray}

The predicted shape of the aging TRM is thus a constant minus a power law of
exponent
$1-x_M$
for the short times ($t\ll t_w$), and on the other hand a simple power law of
exponent $x_1$
at long times ($t\gg t_w$).
In the simple one-level case of all-connected traps with trapping times
distributed
with a
given $x$ (\ref{tau1+x}), one has $x_1=x_M=x$,
which does not yield a realistic TRM-decay shape.
 However, a satisfactory
fit to the TRM measurements can be obtained with only two values of $x$
\cite{jpbdd,jpbev}; one finds
from the initial part of the TRM (or from $\chi''$ measurements) 
$x_M \sim 0.65 - 0.8$ and  from its
late part $x_1=0.05 - 0.35$ (for several samples and different
temperatures). In Fig.7, we show the asymptotic behaviors corresponding to
(\ref{pi1}) for a typical TRM curve, $1-x_M$  being equivalent to the exponent
in (\ref{1-xM}) if $h(t)$ is a power law.



\setcounter{figure}{6}
\begin{figure}
\centerline{\epsfxsize=10cm
\epsffile{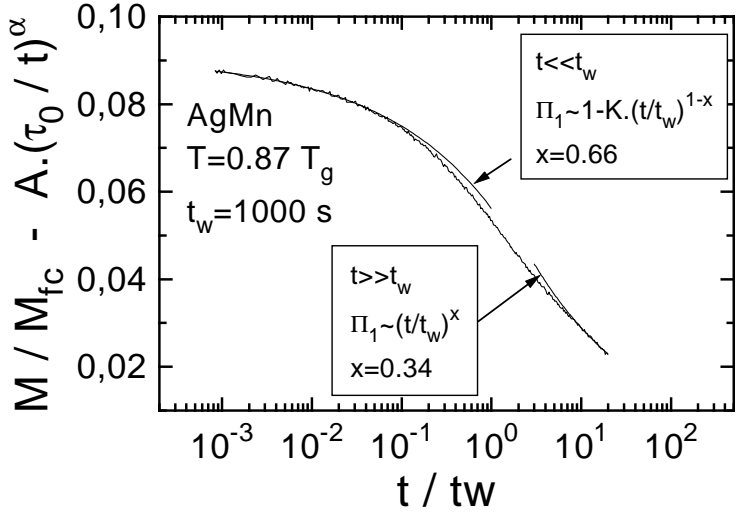}
}
\caption{
Aging part of one of the above TRM curves (Fig. 1 and 3). Both asymptotic
limits $t\ll t_w$ and $t\gg t_w$ have been fitted to (\ref{pi1}). The variation
of
 the effective index $x$ along the different time regimes suggests a more
complex than one-level tree structure.
}
\end{figure}

%

Thus, if one thinks in terms of a one-level picture with no
further assumption, it appears from the shape of the TRM itself
that the effective $x$
which labels the trapping time distribution (\ref{tau1+x}) is {\em different}
in the short and long-time
regimes. Namely, $x$  is closer to 1 for $t\ll t_w$ ($x=x_M$), and decreases
for $t\gg
t_w$ ($x=x_1$). This simple observation
has been translated in terms of a hierarchical organization of the
traps (traps inside traps); the model
of all-connected traps \cite{jpbtrap} (one-level tree) has thus been
generalized in \cite{jpbdd}, of which
we now extract the main points.

The one-level tree is a distribution  of traps of index  say $x=x_1$ ($<1$);
following
the construction of the
Parisi tree of states \cite{beyond}, it can be extended into an n-level tree by
multi\-furcating each trap
into others of index $x_2>x_1$, which themselves subdivide into others, until
some final index $x_n>1$
at the end-branches of the tree. At each level $i$, the $x_i$-distribution of
trapping times corresponds
\cite{beyond,jpbtrap} to an $x_i$-dependent distribution of free energies, and
this yields for the overlap
between the corresponding states a value $q_i(x_i)$ which is the inverse of the
Parisi order parameter
$q(x)$ \cite{beyond}, thus $q_{i+1}>q_i$. Now, the states close to the
end-branches (say the lower part
of the tree)  with $x>1$  can all be visited in finite times, since for $x>1$
the mean value of the distribution
(\ref{tau1+x}) is finite: the corresponding dynamics is {\em stationary}.

Aging phenomena appear when $x<1$, i.e. above a given level $c$ such that
$x_c=1$. The faster
processes of the aging
regime, which are seen in the short-time part of the TRM, occur among the
states which are close to
each other in the hierarchical geometry (large overlap), that is which are
related by a tree-node at a
level very cloes to $c$; therefore they correspond to $x$ close to 1, as is
indeed suggested by the
exponent of the power law behavior at the beginning of the TRM. As time
elapses, more distant states
(of smaller overlap) can be explored, and the corresponding transitions imply
passing  higher nodes
in the tree, of smaller index $x$, in agreement with the smaller exponent of
the power law in the long-time
part of the TRM.

The model can be solved for an arbitrary number of levels; but the
simpler case of a
{\em two-level} tree has been computed and fitted to the TRM experiments
\cite{jpbdd}. The
fits are of a very
good quality for both insulating and metallic samples: for a given $t_w$, the
shape of the decay
curve is quite well reproduced over the whole $t$ scale.
If the number of
metastable states at each
level is infinite, however, the predicted influence of $t_w$ is precisely a
$t/t_w$ scaling (full aging),
as in \cite{jpbtrap}, and this does not satisfactorily account for the
measured $t_w$-dependence, which corresponds to a
sub-aging behavior (see the discussion in Sect. 4).

\subsubsection{Effect of Temperature Changes.}

Now, the effect of the temperature on this tree-like organization of the
metastable states can also be discussed. From the fit of the TRM-data
\cite{jpbev}, the $x_M$-index (e.g. from the short-time part) shows a tendency
to
increase towards 1 as $T$ approaches $T_g$; this suggests that the spin-glass
transition be closer to the REM scenario \cite{REM}, where $x=T/T_g$, than to
the mean-field equilibrium scenario \cite{beyond} where the Parisi parameter
$x$ rather goes to zero at $T_g$.
This leads us to a possible interpretation of the empirical {\em hierarchical
structure versus temperature} in terms of the Parisi-like tree of traps
\cite{jpbdd}. 

$T_g$ is characterized by a certain c-level of the tree where
$x_c=1$; the temperature dependence of $x$ means that, when the temperature is
lowered, this is now a lower (multifurcated) level of the tree which
corresponds to $x=1$ and limits the stationary dynamics. Thus, aging dynamics
restarts among newly born states, as observed in the experiments and
interpreted along the hierarchical scheme of Fig. 6. Conversely, when the
temperature is raised back, the lower aging level comes back to equilibrium,
and aging continues at the upper level, resuming from its previous stage of
evolution. A more detailed comparison between theory and experiment is however
desirable. Note finally that a hierarchical decoupling of time scales has also
been argued on the basis of
``second'' noise spectra by Weissman and collaborators \cite{weiss}.

The influence of temperature is thus crucial in these pictures where activated
processes play the dominant part in aging. Conversely, temperature is somewhat
irrelevant for the aging dynamics of the mean-field models
described above (at least those corresponding to single-scale dynamics). For
example, the effect of temperature variations has been explicitely computed
within the spherical $p=2$ model \cite{Cudd}, with the result that nothing
comparable to experiments happens. In this respect, aging in these single scale
models is again very similar to simple domain growth in a ferromagnet
\cite{Kula}, which is rather insensitive to temperature. It would be very
interesting to understand
how temperature changes affect the dynamics of a multiscale model, such as the
SK model.

\vspace{2cm}
\section{Conclusions}

In this paper, we have presented a survey of the experimental
results concerning
 aging phenomena in spin glasses, focusing on
the description of magnetization relaxation and
low-frequency out-of-phase susceptibility  measurements. We have discussed in
some details how far two theoretical approaches of out-of-equilibrium effects,
namely 
the trap model \cite{jpbtrap,jpbdd,jpbev} and the microscopic approach
\cite{Cuku,Cuku3}, can
succeed in describing different aspects of the dynamics.
\newline
\indent
The relaxation of the out-of-phase susceptibility $\chi"$ towards a non-zero
value indicates the presence of stationary dynamics. This same dynamics can be
found, although less obviously, in the short-time part (compared to the waiting
time $t_w$) of the decay of the thermo-remanent magnetization (TRM). In fact,
an additive combination of a stationary and an aging regime
in the auto-correlation (and thus also in the TRM and in $\chi"$) follows
from the solution of some mean-field models \cite{Cuku,Cuku3,Frme,Cukule,Cudd}.
This striking similarity of mean-field and real 
spin glasses raises the question of the nature of the slow dynamics in these
systems. While one is used to think of thermal activation in mountaneous
landscapes as the source of slow processes, the mean-field models now provide
us with aging phenomena which are due to the flatness of large regions in the
phase space \cite{Kula}, with no crucial role played by the temperature.
\newline
\indent
For the sake of a coherent description of both TRM {\it and}
$\chi"$ data, one should therefore extract the same stationary contribution
from all results, which we have made here in an additive way. The remaining
{\em aging contribution} presents, for the TRM, systematic departures from a
``full aging" situation of a pure $t/t_w$ scaling (``sub-aging")
\cite{aging2,nousaging}. 
When the stationary dynamics of the TRM is additively accounted for,
 the departures from full aging become less pronounced, but still remain.
 For $\chi"$, whose aging regime only overlaps that of the TRM on a limited
range, both full aging and sub-aging scalings remain compatible with the data.
Using some recent analytical developments of the microscopic approach of
spin-glass dynamics \cite{Cuku3}, we have shown that a sub-aging behavior will
appear under some general conditions. Thus, the microscopic theory can now
account for the scaling functions which had been postulated in the past by the
experimentalists on phenomenological grounds \cite{aging2,nousaging}.
\newline
\indent
The experiments have shown that small temperature variations have a strong
effect on aging phenomena \cite{jphys+,uppsaladT,ucladT}. A temperature
decrease restarts the evolution, whereas raising back the temperature lets the
system retrieve its previous stage of aging at this same temperature. These
results have been interpreted in terms of a hierarchical organization of the
metastable states as a function of temperature, in which the valleys of the
free-energy landscape subdivide into others for decreasing temperatures
\cite{jphys+,ucladT}. The mean-field models have not yet brought conclusive
results on this question. The trap model \cite{jpbtrap}, which provides us with
a stochastic picture of aging, has recently been extended \cite{jpbdd} in a way
which sheds some light on the temperature variation experiments.
\newline
\indent
The simple picture of a random walk among all-connected traps (one-level tree
of states) did not take into account the existence of stationary dynamics,
which in this language  is a ``bottom-of-the-traps" dynamics. For this reason,
traps must have an internal structure.
 On the other hand, the TRM shape is related to the index $x$ of the trapping
time distribution. The comparison of the experimental shapes with a one-level
tree model shows that the ``effective $x$" systematically varies along the
curve; it is closer to 1 in the $t\ll t_w$ regime, and smaller in the $t\gg
t_w$ regime \cite{jpbtrap,jpbev,jpbdd}. Thus, the traps which are explored in
the various time regimes are not connected in the same way.
\newline
\indent
These results are fairly well understood within a picture of hierarchically
connected traps, in a multi-level tree geometry where $x$ varies from one level
to the other. The limit value $x=1$ sets the tree level which is at the border
of stationary and non-stationary dynamics. The tendency of $x$ to increase when
the temperature approaches $T_g$ \cite{jpbev} indicates that this border $x=1$
level changes with temperature; the resulting model of a tree-like organization
where the limit of equilibrium dynamics changes with temperature \cite{jpbdd}
is  now very close to the empirical picture of a hierarchy of metastable states
as a function of temperature which had been drawn from the experiments
\cite{jphys+}.
\newline
\indent
This recent development, together with the predictions of slow dynamics and
aging from microscopic models, have created a very stimulating atmosphere. They
let us expect that an even tighter interaction among theoreticians and
experimentalists will bring us closer to a more complete understanding of the
physics of disordered systems.
\newline
\indent
\vspace{5cm}

{\it Acknowledgements}
We would like to thank L. Balents, D. Dean, Vik. Dotsenko, M. Feigel'man,
 J. Kurchan, P. Le Doussal, M. M\'ezard, R. Orbach and G. Parisi for numerous
stimulating discussions all along this work.

\end{document}